%% file: Tonetto_Benhar.tex
\documentclass[prc,twocolumn,nofootinbib,showpacs,amsmath,superscriptaddress,amssymb,floatfix]{revtex4-1}
\usepackage{dcolumn}
\usepackage{bm}
\usepackage{color}
\usepackage{amssymb}
\usepackage{amsmath}
\usepackage{graphicx,epstopdf}
\usepackage{amsfonts}
\usepackage{slashed}
\usepackage{pstricks}
\usepackage{float}
\usepackage{hyperref}
\usepackage{array}
\allowdisplaybreaks
\usepackage{dcolumn}
\usepackage{epsf}
\usepackage{isotope}
\usepackage{comment}
\def\aap{A\&A}

\def\prd{PRD}

\begin{document}
\title{Thermal effects on nuclear matter properties}
\author{Lucas Tonetto}
\affiliation{Dipartimento di Fisica, ``Sapienza'' University of Rome, Piazzale A. Moro, 5. 00185 Roma, Italy}
\affiliation{INFN, Sezione di Roma, Piazzale A. Moro, 5. 00185 Roma, Italy}
%
%\author{Andrea Sabatucci}
%\affiliation{Dipartimento di Fisica, ``Sapienza'' University of Rome, Piazzale A. Moro, 5. 00185 Roma, Italy}
%\affiliation{INFN, Sezione di Roma, Piazzale A. Moro, 5. 00185 Roma, Italy}
%
\author{Omar Benhar}
\affiliation{INFN, Sezione di Roma, Piazzale A. Moro, 5. 00185 Roma, Italy}
\affiliation{Dipartimento di Fisica, ``Sapienza'' University of Rome, Piazzale A. Moro, 5. 00185 Roma, Italy}

\date{\today}

\begin{abstract}
A quantitative description of the properties of hot nuclear matter will be needed for the interpretation of 
the available and forthcoming astrophysical data, providing information  on
the post merger phase of a neutron star coalescence. We have employed a recently developed theoretical model, 
based on a phenomenological nuclear Hamiltonian including two- and three-nucleon potentials, to study the temperature dependence of  
average and single-particle properties of nuclear matter relevant to astrophysical applications. 
The possibility to represent the results of microscopic calculations using simple and yet 
physically motivated parametrisations of thermal effects, suitable for use in numerical simulations of astrophysical processes, is  
also discussed.
\end{abstract} 

%\keywords{ Dense matter -- Equation of state -- Stars: neutrons -- Gravitational waves}
%{\pacs{XXXX, YYYY, ZZZZ}

\index{}\maketitle

%----------------------------------------
\section{Introduction}
%----------------------------------------

Understanding the structure
and dynamics of hot nuclear matter at microscopic level is long known to be essential 
for the description of both supernov\ae  \ and proto neutron stars~\citep{burrows1986,keil1995,pons1999,camelio2017}. 
More recently, thermal modifications of the equation of state (EOS) of neutron star matter have been also shown to play a critical role
in the merger and postmerger phases of binary neutron star coalescence~\citep{baiotti2017,raithel2021,figura2020,figura2021,hammond2021}. 
In this context, it has to be pointed out that an accurate 
description of finite-temperature effects is needed to study not only the equilibrium properties determining the density dependence of matter pressure, 
but also the occurrence of phenomena involving dissipation mechanisms, such as bulk viscosity \citep{alford2018} and neutrino emission \citep{camelio2017}.

The large data set of zero-temperature EOSs available for use in simulations and data analysis\textemdash   for a comprehensive catalogue see Ref.~\cite{CompOSE}\textemdash is contrasted by a scarce number of EOSs of hot nuclear matter spanning the 
relevant regime, which is believed to extend to temperatures as high as 100 MeV~\citep{raithel2019,oechslin2007}. 

The EOS of hot nuclear matter is often obtained using Skyrme-type effective interactions~\cite{Prakash:1997} or 
the Relativistic Mean Field (RMF) approach~\citep{Kaplan:2014}. More comprehensive studies of the properties of  
neutron star matter at nonzero temperature have been carried out within the framework of Nuclear Many-Body Theory, in which nuclear
dynamics is described by a phenomenological Hamiltonian, strongly constrained by the observed properties of the two- and
three-nucleon systems~\cite{PhysRevC.100.054335}. Recent calculations along this line have been performed using both $G$-matrix perturbation theory~\cite{PhysRevC.100.054335} and the formalism of Correlated Basis Functions~\cite{benhar2022}.

The authors of Refs.~\cite{BL2017,benhar2022} have developed a procedure to obtain from a phenomenological nuclear Hamiltonian 
a well-behaved effective potential, suitable to carry out perturbative calculations in the basis of eigenstates of the  non interacting system.
This approach, in which the effects of irreducible three-nucleon interactions are consistently taken into account at microscopic level, allows 
to perform calculations of a variety of properties of dense nuclear matter with arbitrary proton fraction and
temperatures in the region of $T \ll m_\pi$, $m_\pi \approx 150$ MeV being the mass of the $\pi$-meson, in which thermal effects  
are not expected to significantly affect strong-interaction dynamics. Thermodynamic consistency is also achieved by
construction, through a proper definition of the  grand canonical potential~\cite{benhar2022}.

The present work is primarily meant as a follow-up to the study of~\citet{benhar2022}, and provides a detailed analysis 
of the impact of thermal effects on specific properties of charge-neutral and $\beta$-stable matter relevant to neutron stars, 
such as the proton and neutron energy spectra and effective masses. 

We will also examine the possibility of using simple approximated procedures to parametrise deviations from  
the zero-temperature EOS associated with thermal effects. The development of such procedures is of utmost importance, 
because their availability will enable to perform numerical simulations using EOSs based on 
on a reliable treatment of the zero-temperature limit. Pinning down the validity and limitations of the proposed 
procedures, through a direct comparison with the predictions of fully microscopic calculations, will help 
to firmly establish their applicability. 

A widely used, although admittedly oversimplified, parametrisation
is obtained from the so-called ``hybrid-EOS" approach, in which thermal modifications of the thermodynamic functions of cold 
nuclear matter are 
approximated by the corresponding quantities of an ideal fluid~\citep{bauswein2010,figura2020,hotokezaka2011,endrizzi2016,dietrich2017a,dietrich2017b}. 

Within this scheme, pressure and specific internal energy are respectively written in the form 
\begin{align}
\nonumber
p & = p_{\rm cold} + p_{\rm th}  \ , \\
\nonumber
e & = e_{\rm cold} + e_{\rm th} \ , 
\end{align}
and the thermal contribution to the pressure at matter density $\varrho$ and temperature $T$ is parametrised by the adiabatic 
index, $\Gamma_{\rm th}$, according to 
\begin{align}
\label{hybrid}
p_{\rm th}(\varrho,T) = \varrho~e_{\rm th} (\Gamma_{\rm th} - 1) \ . 
\end{align}

The above procedure involves the drastic assumption that the adiabatic index be independent of both density and temperature.
However, a comparison between the pressure obtained from Eq.~\eqref{hybrid} and that resulting from microscopic 
calculations based on advanced models of nuclear dynamics shows that  $\Gamma_{\rm th}$ does, in fact, depend strongly 
on density, and that the dependence on temperature, while being weaker, is also non negligible~\cite{figura2020}. 

A more advanced parametrisation, aimed at improving the description of the thermal pressure in the high-density region, has been recently proposed 
by~\citet{raithel2019}. Within this approach, the prediction of the ideal fluid model\textemdash which is known to overestimate pressure at large densities\textemdash is replaced with that obtained from the leading term of the Sommerfeld expansion, which allows to systematically include degeneracy effects~\cite{Constantinou}. Microscopic nuclear dynamics is taken into account, using nucleon effective masses obtained from RMF models of nuclear matter.

To assess the accuracy and range of applicability of the simple and yet physically motivated parametrisation of Ref.~\cite{raithel2019}, we have compared its predictions to the results obtained from microscopic calculations of $\beta$-stable matter at temperatures up to $50$ MeV,  carried out within the formalism described in Ref.~\cite{benhar2022}.

The manuscript is organised as follows. In Sect.~\ref{sec:formalism} we outline the dynamical model underlying our theoretical approach, as well as the 
main elements of the formalism employed to study the properties of hot nuclear matter. \ldots\ldots

%----------------------------------------
%\section{Properties of hot nuclear matter} \label{sec:hot_matt}
%----------------------------------------
%----------------------------------------
\section{Theoretical model} \label{sec:formalism}
%----------------------------------------

In this section, we summarise the main features of our theoretical model. 
We discuss both the the underlying description of nuclear dynamics and the 
formalism employed to carry out calculations of the relevant properties of hot nuclear matter.

\subsection{Nuclear dynamics} \label{subsec:nucdyn}

Nuclear Many-Body Theory (NMBT) is based on the hypothesis that all nucleon systems\textemdash from the deuteron to neutron stars\textemdash  can be described in terms of point like protons and neutrons, whose dynamics is dictated 
by the Hamiltonian%
\begin{equation}
\label{hamiltonian}
H=\sum_{i}\frac{p_i^2}{2m} + \sum_{i<j}v_{ij}+\sum_{i<j<k}V_{ijk} \ ,
\end{equation}
with $m$ and ${\bf p}_i$ denoting mass and momentum of the $i$-th particle.\footnote{In this article, we adopt the 
system of natural units, in which $\hbar=c=k_B=1$, and, unless 
otherwise specified, neglect the small proton-neutron mass difference.}

The nucleon-nucleon (NN) potential, usually written in 
the form
\begin{equation}
    v_{ij} = \sum_p v^p (r_{ij}) O_{ij}^p \ ,
    \label{eq:vij}
\end{equation}
where $r_{ij} = |{\bf r}_i - {\bf r}_j|$ is the distance between the interacting particles, is designed  
to reproduce the measured properties of the two-nucleon system, in both bound and scattering states, and reduces 
to the Yukawa one-pion exchange potential at large distances. The sum in Eq.~\eqref{eq:vij} includes up 
to eighteen terms, the corresponding operators, $O^p$, being needed to describe the strong spin-isospin 
dependence and non central nature of nuclear forces, as well as the occurrence spin-orbit interactions and 
small violations of charge 
symmetry and charge independence~\cite{AV18}.

The addition of the three-nucleon (NNN) potential $V_{ijk}$ is needed to take into account the effects of irreducible three-body 
interactions, reflecting the occurrence of processes involving the internal structure of the nucleons. 

%Phenomenological 
%models of $V_{ijk}$ are constructed in such a way as to explain the observed binding energies of \isotope[3][]{He} and
%\isotope[4][]{He}, as well as the saturation density of isospin symmetric mater inferred from nuclear systematics.

The results reported in this article have been obtained using an {\it effective interaction} derived from the phenomenological 
Hamiltonian comprising the Argonne $v_6^\prime$ (AV6P) NN potential~\cite{AV6P} and the Urbana IX (UIX) NNN potential~\cite{UIX_1,UIX_2}.

The AV6P potential is determined projecting the full Argonne $v_{18}$ potential of Ref.~\cite{AV18} (AV18) onto the operator basis comprising  
the terms with $p \leq 6$ in the right hand side of Eq.~\eqref{eq:vij}. It predicts the binding energy and electric quadrupole 
moment of the deuteron with accuracy of 1\%, and 4\%, respectively, and provides an excellent fit of the NN scattering phase 
shifts in the $^1{\rm S}_0$ channel, corresponding to total spin and isospin $S=0$ and $T=1$, and relative angular momentum $\ell = 0$.

The UIX potential is written in the form
\begin{align}
V_{ijk}=V_{ijk}^{2\pi}+V_{ijk}^{R} \ ,
\end{align}
where the first term is the attractive Fujita-Miyazawa potential\textemdash describing two-pion exchange NNN interactions 
with excitation of a $\Delta$-resonance in the intermediate state\textemdash while 
$V_{ijk}^{R}$ is a purely phenomenological repulsive term. The strength of 
$V_{ijk}^{2\pi}$, is adjusted to explain the observed ground-state energies of
\isotope[3][]{He} and \isotope[4][]{He}, while that of the isoscalar repulsive contribution 
is fixed in such a way as to reproduce the saturation density of isospin symmetric matter, inferred from nuclear systematics.

Recent studies of the EOS of cold neutron matter\textemdash performed by~\citet{Lovato:2022} using state-of-the-art
computational techniques\textemdash show that the
predictions of the somewhat simplified AV6P+UIX Hamiltonian are very close to those obtained from the full AV18+UIX model, providing the basis of the widely employed EOS of Akmal, Pandharipande and Ravenhall~\cite{Akmal:1997,Akmal:1998}.

The procedure to derive the effective interaction, thoroughly described 
in Refs.\cite{lovato2013,eos0,benhar2022}, exploits the formalism of correlated basis functions (CBF) and cluster 
expansion techniques to take into account the effects of strong nucleon-nucleon correlations, arising from the presence of a strong 
repulsive core in the NN potential.
The resulting density-dependent effective potential\textemdash which can be written as in Eq.~\eqref{eq:vij} with the sum 
in the right-hand side limited to $p \leq6$\textemdash is well behaved, and consistently includes the 
contributions of NN and NNN interactions. As a consequence, it is expected to be well suited to perform perturbative calculations of nuclear 
matter properties in the density regime relevant to neutron stars. 
 
%-----------------------------------------------------------------------
\subsection{Perturbation theory at finite temperature }
%-----------------------------------------------------------------------

At first order in the CBF effective interaction $v^{\rm eff}$, the internal energy per nucleon of nuclear matter at 
baryon density $\varrho$, temperature $T$, and proton fraction $Y_p$ can be written in the form~\cite{benhar2022}
\begin{widetext}
\begin{align}
\label{int:en}
\frac{E}{N} = \frac{1}{N} \Big\{ \sum_{ \alpha {\bf k} }   \ \frac{ {\bf k}^2 }{2m}  \ n_\alpha(k,T)  
+ \frac{1}{2} \sum_{ \alpha {\bf k} } \sum_{ \alpha^\prime {\bf k}^\prime } 
\langle \alpha {k} , \alpha^\prime  {k}^\prime | v^{\rm eff} | \alpha {k} , \alpha^\prime  {k}^\prime \rangle_A  
\ n_\alpha(k,T) n_{\alpha^\prime}(k^\prime,T) \Big\} \ .
\end{align} 
\end{widetext}
In the above equations, the index $\alpha = n, p$ labels neutrons and protons, respectively, ${\bf k}$ is the nucleon momentum, $k = |{\bf k}|$, 
and $| \alpha {k} , \alpha^\prime  {k}^\prime \rangle_A$ denotes an antisymmetrised two-nucleon state. Note that conservation of baryon number
requires that $Y_n = 1 - Y_p$.

The temperature dependence is 
described by the Fermi distribution 
\begin{align}
\label{fermidist}
n_\alpha(k,T)  = \Big\{ 1 +  \exp { [ \beta ( e_{\alpha k} - \mu_\alpha ) ] } \Big\}^{-1}  \ .
\end{align}
where the single-particle energy is defined as, 
\begin{align}
\label{ek}
e_{\alpha k} = e^{\rm HF}_{\alpha k} + \delta e \ ,
\end{align}
with
\begin{align}
\label{eHFk}
e^{\rm HF}_{\alpha k} = \frac{{\bf k}^2}{2m} & + \sum_{ \alpha^\prime {\bf k}^\prime } 
\langle \alpha {k} , \alpha^\prime  {k}^\prime | v^{\rm eff} | \alpha {k} , \alpha^\prime  {k}^\prime \rangle_A  \ n_\alpha(k^\prime,T) \ ,
\end{align}
and
\begin{align}
\nonumber
 \delta e = \frac{\varrho}{2} \sum_{ \alpha {\bf k} } \sum_{ \alpha^\prime {\bf k}^\prime } & 
 \langle \alpha {k} , \alpha^\prime  {k}^\prime | \frac{\partial v^{\rm eff}}{\partial \varrho} | \alpha {k} , \alpha^\prime  {k}^\prime \rangle_A \\
\label{deltae}
 & \times  n_\alpha(k,T) n_{\alpha^\prime}(k^\prime,T) , 
\end{align}
The correction to the Hartree Fock (HF) spectrum is needed to satisfy 
the requirement of thermodynamic consistence, and vanishes in the case of a density-independent potential; see Ref.~\cite{benhar2022} for details. 

The chemical potentials $\mu_\alpha$ are determined by the normalisation conditions 
\begin{align}
\label{def:chempot}
\frac{2}{V} \sum_{ \alpha {\bf k} } n_\alpha(k,T)  = \varrho_\alpha \ , 
\end{align}
where $V$ is the normalisation volume, and the number density of  particles of species $\alpha$ is defined as $\varrho_\alpha = Y_\alpha \varrho$. Note that the above definitions 
imply that both the single-nucleon energies and the chemical potentials depend on temperature through the Fermi distribution.

The entropy per nucleon is also defined in terms of the distribution of Eq.~\eqref{fermidist} as
\begin{align}
\label{def:S}
\frac{S}{N}   = &- \frac{1}{N} \sum_{ \alpha {\bf k} }  
\Big\{  n_\alpha(k,T) \ln {n_\alpha(k,T)} \\
\nonumber
& +  \big[ 1 - n_\alpha(k,T) \big]  \ln{ \big[ 1 - n_\alpha(k,T) \big] } \Big\} \ .
\end{align}
Finally, the Helmoltz free energy per nucleon is obtained combining Eqs.~\eqref{int:en} and~\eqref{def:S} 
in the forrm
\begin{align}
\label{def:free}
\frac{F}{N} = \frac{1}{N} \big( E - T S \big) \ .
\end{align}

%----------------------------------------
\section{Thermal effects} \label{sec:hot_matt}
%----------------------------------------

In the temperature regime considered in the present study,  thermal modifications of nuclear matter properties 
arise primarily from the Fermi distribution, defined by Eq,\eqref{fermidist}. Comparison to the $T \to 0$ limit
\begin{align}
\label{dist:0}
n_\alpha(k,0) = \theta(\mu_\alpha-e{_{\alpha k}}) \ , 
\end{align}
where $\theta(x)$ is the Heaviside theta-function, shows that the probability distribution $n_\alpha(k,T>0)$ is reduced from unity  
in the region corresponding to $\mu_{\alpha}  -T \lesssim  e{_{\alpha k}} \lesssim \mu_{\alpha}$, while acquiring a non vanishing positive value for
$\mu_{\alpha}  \lesssim e{_{\alpha k}} \lesssim  \mu_{\alpha} + T$. It follows that, for any given temperature $T$, the extent of thermal modifications to the 
Fermi distribution is driven by the ratio $2T/\mu_{\alpha k}$. This observation in turn implies that, because the chemical potential is a monotonically increasing function of  the particle density $\varrho_\alpha$ over a broad range of temperatures, for any given $T$ thermal effects are more significant at lower $\varrho_\alpha$. On the other hand,  they become vanishingly small in the high-density regime, in which degeneracy becomes dominant.  
 
The density-dependence of thermal effects\textemdash that also affects the particle energies and chemical potentials, defined by 
Eqs.~\eqref{ek} and~\eqref{def:chempot}, respectively\textemdash plays a significant role in the determination of the properties of multicomponent systems, such as 
charge-neutral $\beta$-stable matter, in which different particles have different densities.

%----------------------------------------
\subsection{Composition of charge-neutral $\beta$-stable matter}
\label{composition}
%----------------------------------------

In charge-neutral matter consisting of neutrons, protons and leptons in equilibrium with respect to the weak interaction processes 
\begin{align}
n \to p + \ell + {\bar \nu}_\ell \ \ \ \ , \ \ \ \ p + \ell^- \to n + \nu_\ell \ , 
\end{align} 
where $\ell = e, \mu$ labels the lepton flavour, the proton fraction $Y_p = \varrho_p/\varrho_n$ is uniquely determined by the equations
\begin{align} 
\label{beta:eq1}
\mu_n - \mu_p = \mu_{\ell} \ ,
\end{align}
\begin{align} 
\label{beta:eq2}
Y_p = \sum_\ell Y_\ell \ .
\end{align}
At densities such that the electron chemical potential does not exceed the rest mass of the muon, $m_\mu =~105.7$~MeV, the sum appearing in the 
above equation includes electrons only. However, at higher densities\textemdash typically at $\varrho \gtrsim \varrho_0$, with $\varrho_0 = 0.16 \ {\rm fm}^{-3}$ being the equilibrium density 
of isospin-symmetric matter\textemdash  the appearance of muons becomes energetically favoured, and must be taken into account.  

%%%%%%%%%%%%%%%%%%%%
\begin{figure}[htb]
%\vspace{0.5cm}
\includegraphics[scale=0.6]{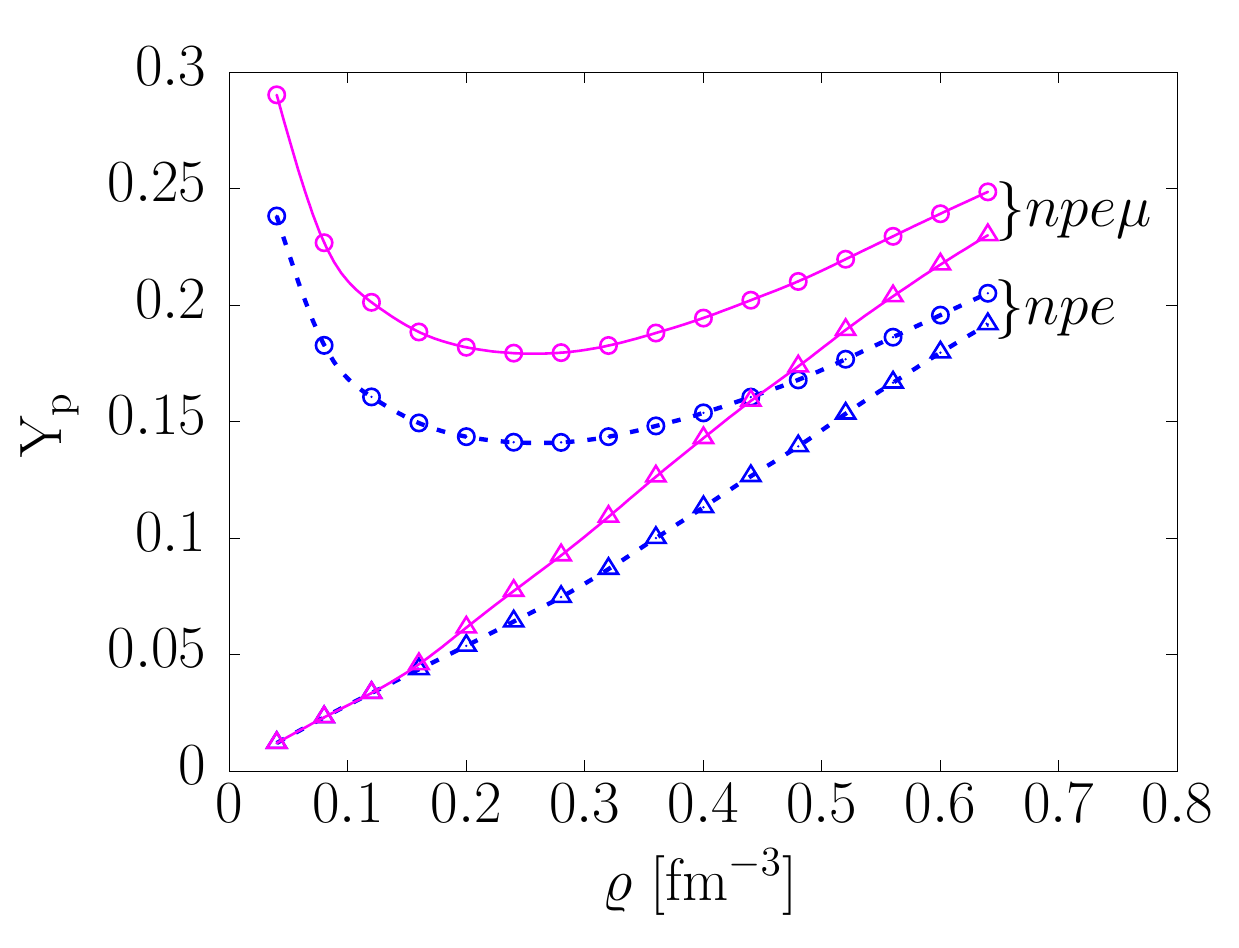}
\vspace{-0.25cm}
     \caption{Density dependence of the proton fraction in charge-neutral $\beta$-stable matter. Solid lines marked with triangles and circles correspond to $npe\mu$ matter at $T=$ 0 amd 50 MeV, respectively.
     The same quantities in $npe$ matter are represented by dashed lines.}
\label{prot:frac}
\end{figure}
%%%%%%%%%%%%%%%%%%%%

The solid lines of Fig.~\ref{prot:frac} show the density dependence of the proton fractions corresponding to $\beta$-equilibrium of matter consisting of protons, neutrons, electrons and muons, or $npe\mu$ matter, at $T=$ 0 (triangles) and 50 MeV (circles); all results have been obtained using the formalism described in Ref.~\cite{benhar2022}. For comparison, the same quantities in $npe$ matter, in which the muon contribution is not included, are displayed 
by the dashed lines. 

The most prominent thermal effect is a significant departure from the monotonic behaviour observed in cold matter. The emergence of a minimum in the 
density dependence of the proton fraction results from the interplay between the thermal and degeneracy contributions to the  chemical potentials appearing in Eq.~\eqref{beta:eq1}. 
For $T \gtrsim 20$~MeV and low density, typically $\varrho \lesssim \varrho_0$, the thermal contribution\textemdash whose leading order term can be written in the form $\delta \mu_\alpha \propto T^2/\varrho_\alpha^{2/3}$\textemdash turns out to be much larger for protons than for neutrons, and $\beta$-equilibrium requires large proton fractions.

%----------------------------------------
\subsection{Fermi distributions}
%----------------------------------------

The Fermi distribution of Eq.~\eqref{fermidist} depends on temperature both explicitly, through the factor $\beta = 1/T$ appearing in the argument of the exponential, and implicitly, through the $T$-dependence of both $e{_{\alpha k}}$ and $\mu_\alpha$. Because the calculation of single-particle energies and chemical potentials in turn involves the Fermi distribution, $e{_{\alpha k}}$, $\mu_\alpha$ and $n_\alpha(k,T)$ must, in fact, be determined self-consistently, appliying an iterative procedure.
 
\begin{figure}[htb]
\vspace{0.25cm}
\includegraphics[scale=0.6]{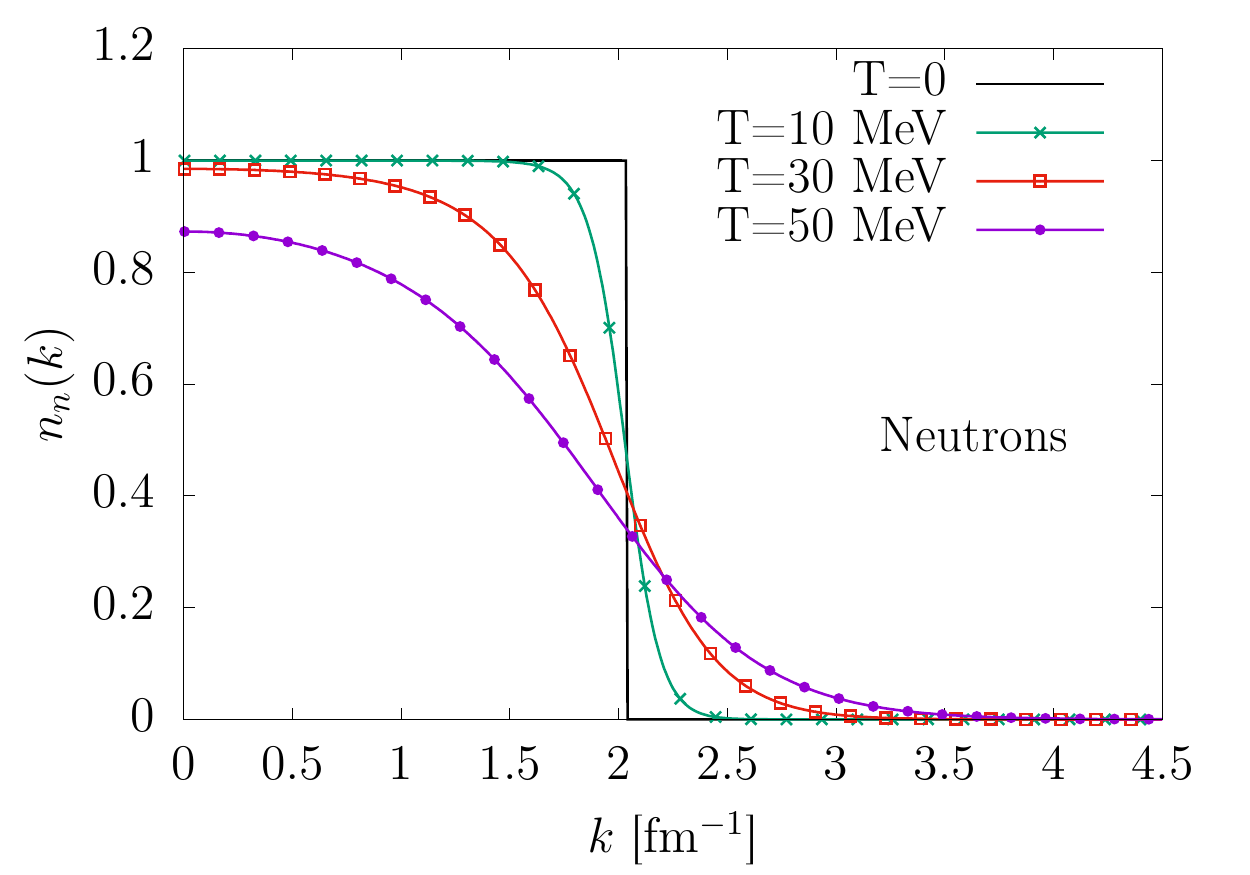}
\includegraphics[scale=0.6]{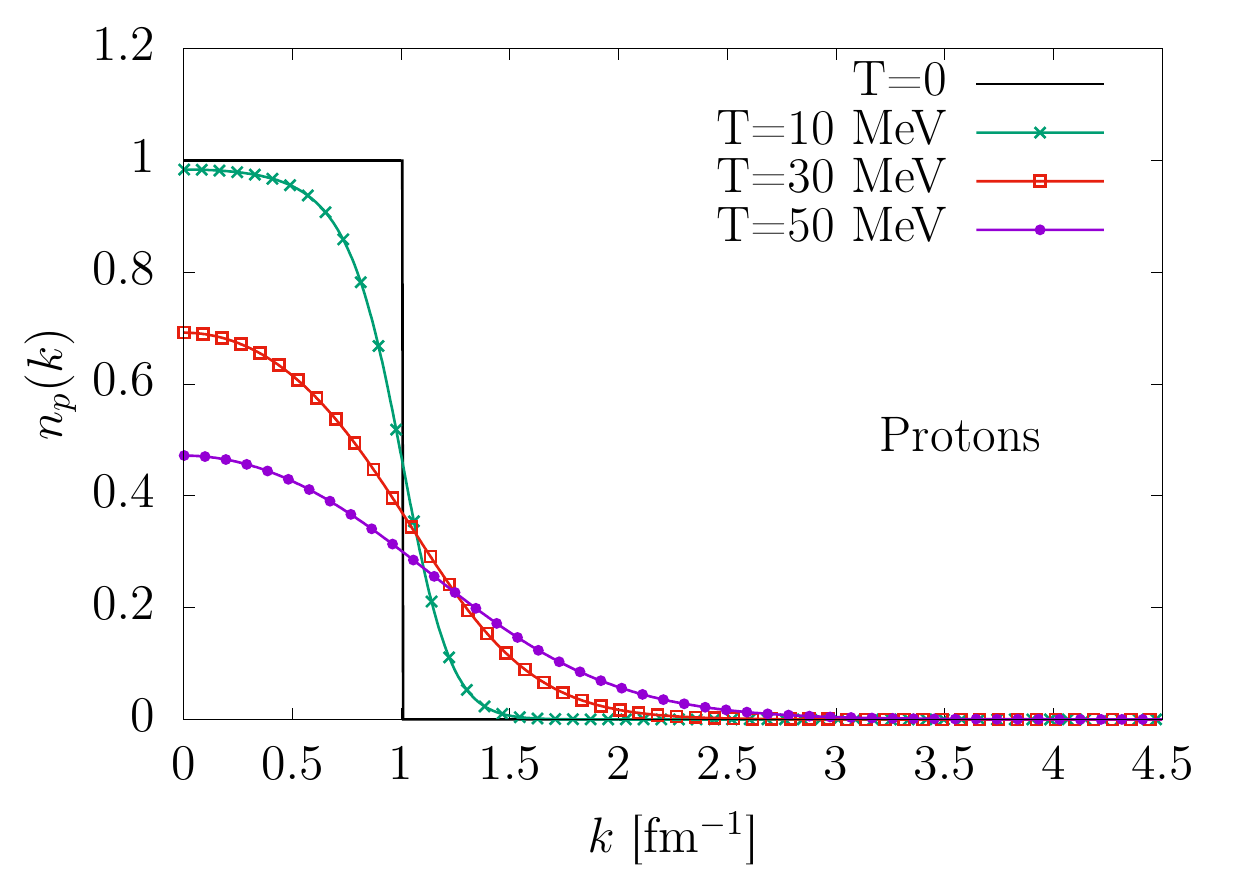}        
\caption{Neutron and proton Fermi distributions in charge-neutral $\beta$-stable $npe\mu$ matter at baryon density $\varrho=0.32 \ \mathrm{fm}^{-3}$.}
\label{fig:fermidist}
\end{figure}

Figure~\ref{fig:fermidist} shows the distributions of neutrons and protons in charge-neutral $\beta$-stable $npe\mu$ matter at baryon density $\varrho=0.32 \ \mathrm{fm}^{-3}$.

%%%%%%%%%%%%%%%%%%%%
%%%%%%%%%%%%%%%%%%%%

It is apparent that, as pointed out in the previous section, thermal modifications to $n_\alpha(k,T)$\textemdash extending over a region of width $2 T$ around the Fermi momentum $k_{F\alpha} = (3 \pi^2 \varrho_\alpha)^{1/3}$\textemdash  depend on  {\it both} temperature {\it and} density.
As a consequence, for any given temperature $T$ they are more pronounced in the case of protons, whose density is suppressed by a factor ${\rm Y}_p/(1-{\rm Y}_p) \ll 1$
with respect to the neutron density.

%----------------------------------------
\subsection{Nucleon energy spectra and effective masses}
%----------------------------------------

\begin{figure}[h]
%\vspace{0.5cm}
\includegraphics[scale=0.6]{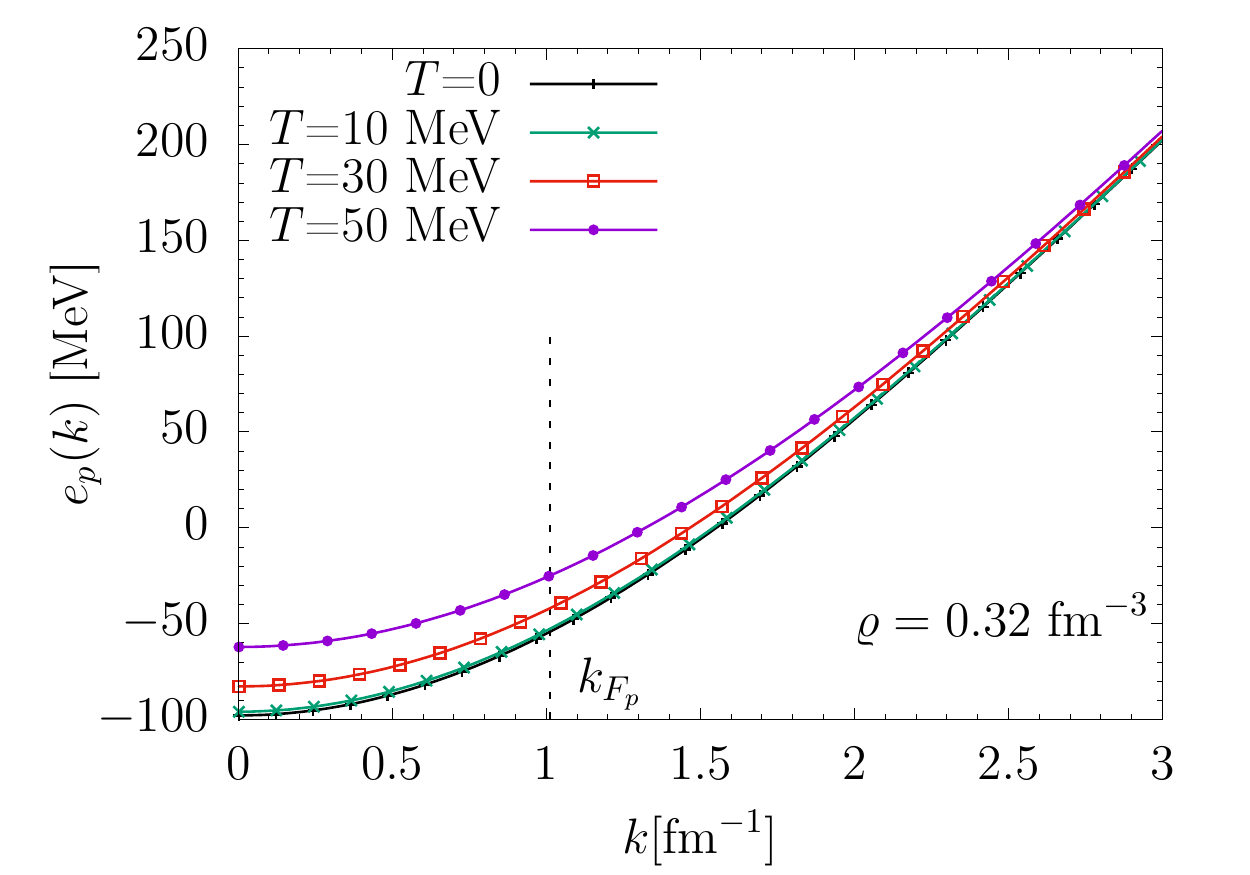}
\includegraphics[scale=0.6]{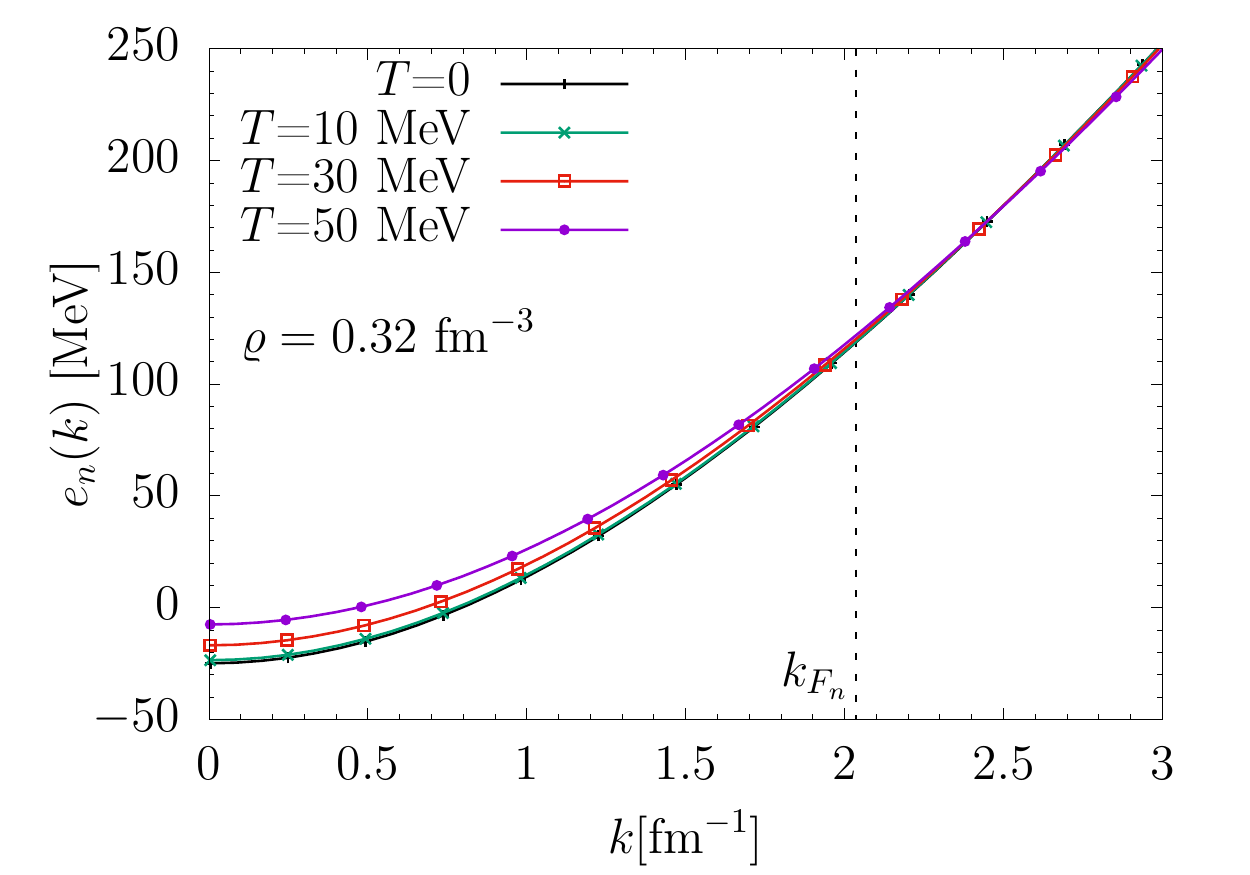}       
\caption{Neutron and proton spectra in $\beta$-stable $npe\mu$ matter at baryon density $\rho=0.32 \ \mathrm{fm^{-3}}$, and temperatures in the range $0 \leq T \leq 50$~MeV.}
\label{fig:spectrum2n0}
\end{figure}

The proton and neutron spectra employed to calculate the Fermi distributions of Fig.~\ref{fig:fermidist}\textemdash coresponding to $\beta$-stable $npe\mu$ matter at baryon density $\varrho = 2 \varrho_0$\textemdash are displayed in Fig.~\ref{fig:spectrum2n0}.
It is apparent that $e{_{\alpha k}}$ is an increasing
function of temperature at all values of $k$, with the $T$-dependence being stronger at lower momentum. At $k=0$ the difference between the energies corresponding to $T=0$ and 50 Mev reaches $\sim 35.8$~MeV for protons, and $\sim 17.5$~MeV for neutrons. In the case of protons, a $\sim 29$ MeV increase with respect to the
zero-temperature spectrum is still clearly visible at $k = k_{F_p}$,  $k_{F_p} = 1.01 \ {\rm fm}^{-1}$ being the proton Fermi momentum, while the
$T=0$ and 50 MeV neutron spectra at  $k = k_{F_n}$,  with $k_{F_n} = 2.04 \ {\rm fm}^{-1}$, are nearly indistinguishable.

%%%%%%%%%%%%%%%%%%%%
%%%%%%%%%%%%%%%%%%%%

In theoretical calculations of nuclear matter properties of astrophysical interest\textemdash such as the neutrino emission rates~\citep{camelio2017}, 
and the shear and bulk viscosity coefficients~\citep{benharvalli2007,alford2018,alford2021}\textemdash the relevant information comprised in proton and 
neutron spectra is captured by the corresponding effective masses $m_\alpha^\star$, defined by the equations
\begin{align}
\label{def:mstar}
\frac{1}{m_\alpha^\star} = \left(  \frac{1}{k} \frac{ d e_{\alpha k} }{ d k }\right)_{k = k_{F_\alpha}} \ .
\end{align}
The role played by the effective masses can be readily grasped considering that they determine the dispersion
relations of matter constituents, which in turn affect their collision rates through both the incident flux and 
the available phase space.

\begin{figure}[h]
\includegraphics[scale=0.6]{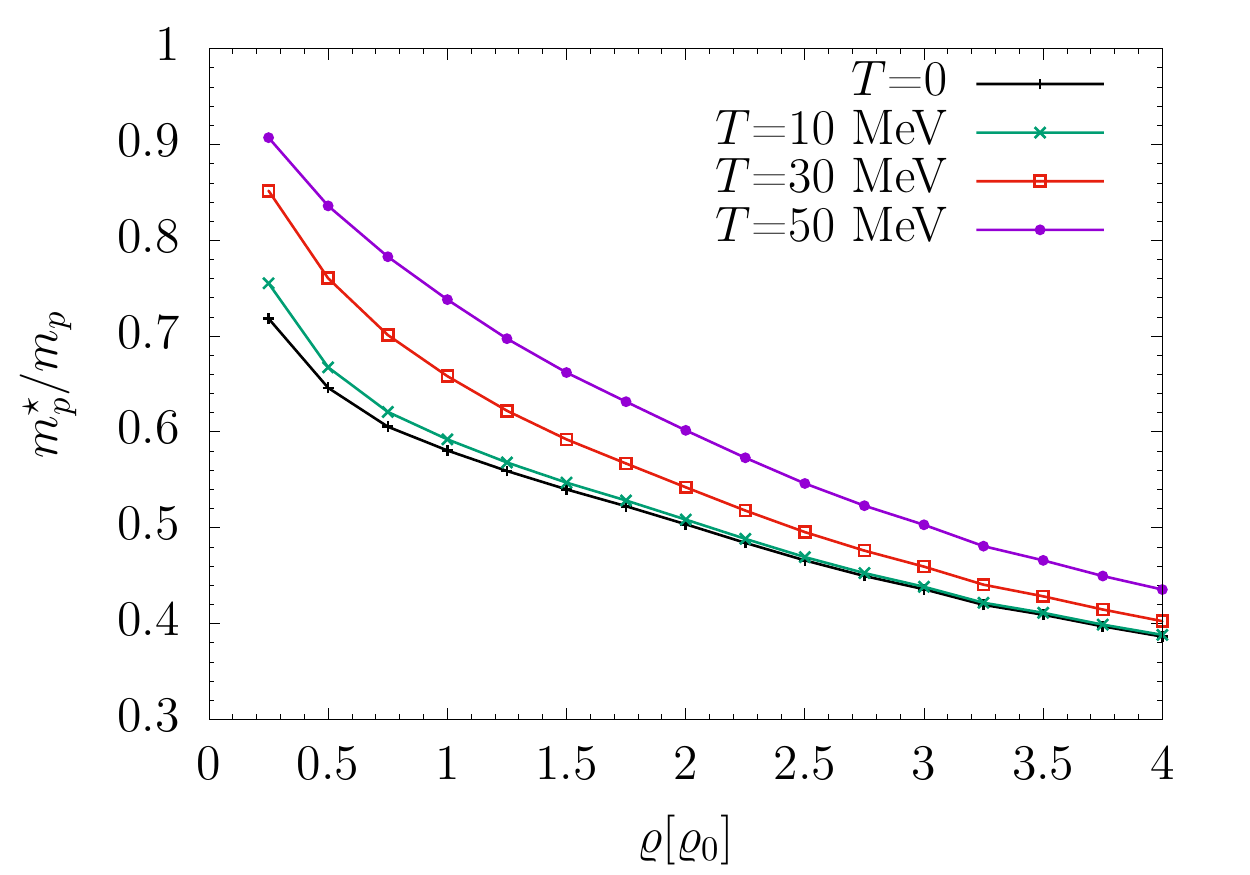}
\includegraphics[scale=0.6]{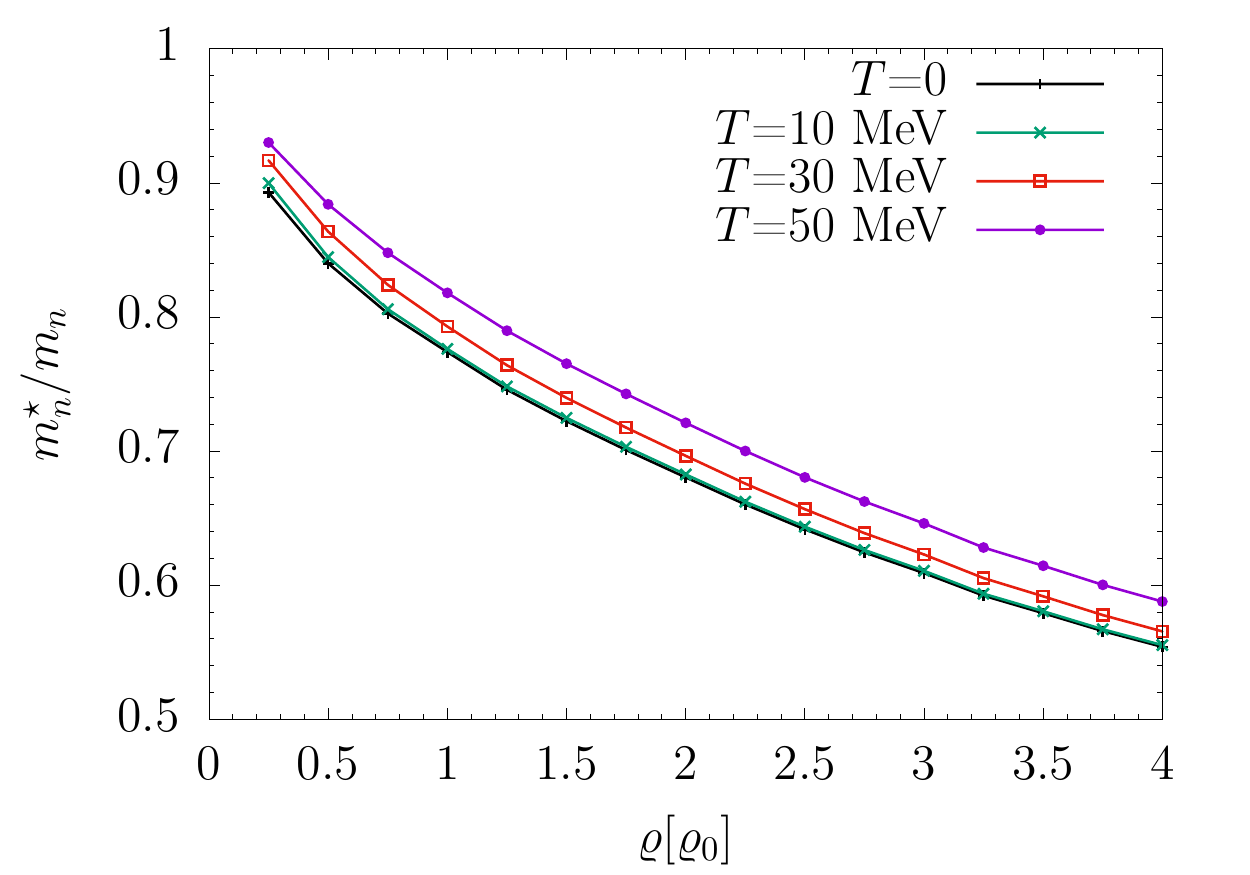}
%vspace*{-.1in}
\caption{Density dependence of the proton and neutron effective masses of charge-neutral $\beta$-stable matter at temperature $0\leq T \leq 50$~MeV.
Baryon densities are measured in units of the equilibrium density of cold isospin-symmetric matter.}
\label{fig:effmass}
\end{figure}

The density dependence of the proton and neutron effective masses of charge-neutral $\beta$-stable $npe\mu$ matter at temperature $0\leq T \leq 50$~MeV is 
illustrated in Fig.~\ref{fig:effmass}. It clearly appears that, regardless of temperature,  ${m_\alpha^\star}$ is a monotonically decreasing function of baryon density. This behaviour is consistent with the results of calculations carried out within the RMF approach, recently
reported in Ref.~\cite{raithel2021}. 
For neutrons, thermal effects\textemdash measured by the departure from the zero-temperature effective mass\textemdash turn out to be limited to $\sim 5$\% over the whole temperature and density 
range considered. For protons, on the other hand, their size for $T=50$ MeV turns out to be $\sim 25$\% at $\varrho = \varrho_0$, 
and is still $\gtrsim 10$\% at $\varrho = 4\varrho_0$.

%%%%%%%%%%%%%%%%%%%%
%%%%%%%%%%%%%%%%%%%%

The nucleon effective masses are routinely used to parametrise the momentum dependence of the nucleon spectra in cold nuclear matter 
according to~\cite{camelio2017}
\begin{align}
\label{ek:quad}
e_{\alpha k} = \frac{k^2}{2 m^\star_0} + U_\alpha \ , 
\end{align}
where $m^\star_0$ denotes the value of  ${m_\alpha^\star}$  at $T=0$, while the offset $U_\alpha$ is determined by the requirement that the 
above approximation reproduce the spectrum obtained from the full microscopic 
calculation in the $k~\to~0$ limit.

%%%%%%%%%%%%%%%%%%%%
\begin{figure}[h]
%\vspace{0.5cm}
\includegraphics[scale=0.55]{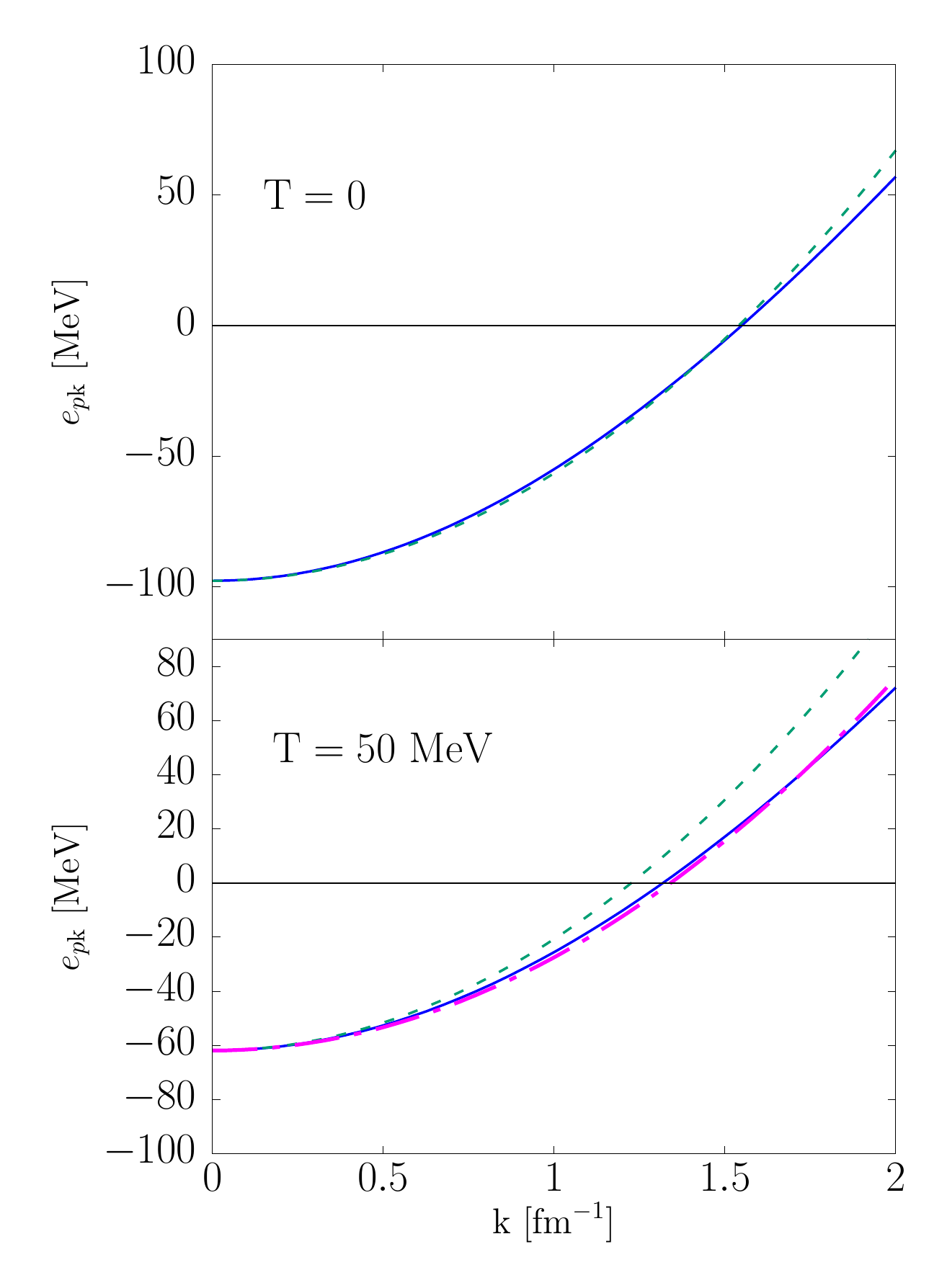} 
\caption{Proton spectra in charge-neutral, $\beta$-stable matter at $T=0$ (upper panel) and 50 MeV (lower panel). The solid lines represent 
results of calculations carried out using Eqs.~\eqref{ek}-\eqref{deltae}, while the dashed lines have been obtained from the 
quadratic approximation of Eq.~\eqref{ek:quad} with the zero-temperature effective mass. The dot-dash line in the lower panel 
illustrates the effect of the thermal dependence of $m_p^\star$; see text for details. }
\label{quadratic:spectra}
\end{figure}
%%%%%%%%%%%%%%%%%%%%

In Fig.~\ref{quadratic:spectra} the proton spectra in $\beta$-stable $npe\mu$ matter at baryon density $\varrho$ = 0.32 fm$^{-3}$ and 
temperature $T=0$ and 50~MeV, obtained from Eqs.~\eqref{ek}-\eqref{deltae}, are 
compared to those computed using Eq.~\eqref{ek:quad}. At $T=0$ the quadratic approximation turns out to be remarkably accurate 
up to momenta largely above the Fermi momentum, $k_{F_p}~=~1.01 \ {\rm fm}^{-1}$. At $T=50$ MeV, on the other hand, 
the agreement between the results of the two calculations is somewhat degraded; the discrepancy is $\sim 25$\% at $k =  k_{F_p}$, and monotonically increases with $k$. 

The spectra displayed in the bottom panel of Fig.~\ref{quadratic:spectra} clearly show  that the accuracy of Eq.~\eqref{quadratic:spectra} at  $T > 0$ can be significantly improved by taking into account the 
temperature dependence of the effective mass, which amounts to replacing $m^\star_0$ with the appropriate finte-temperature 
value, obtained from Eq.~\eqref{def:mstar}.

In the literature, the temperature dependence of $e{_{\alpha k}}$ is often disregarded, and the 
properties of nuclear matter at $T>0$ are calculated using zero-temperature spectra. 
This approximation, referred to as {\it Frozen Correlations Approximation} (FCA), has been recently employed 
in the studies of binary neutron star mergers of~\citet{figura2020,figura2021}. 
The results reported in Ref.~\cite{baldo1999} suggest that the FCA has a nearly negligible effect on the thermodynamic 
properties of nuclear matter at $T \lesssim 30 \ \mathrm{MeV}$. However, its accuracy has been 
shown to deteriorate at larger temperatures~\cite{BS:AA}.
The validity of the assumption underlying the FCA can be gauged Figs.~\ref{fig:spectrum2n0}-\ref{fig:spectrum3n0}.
The implications of using this approximation scheme in calculations of nuclear matter properties will be discussed further 
in the next section. 
 
%----------------------------------------
\subsection{Chemical potentials and matter composition}
%----------------------------------------

\begin{figure}[h]
%\vspace{0.5cm}
\includegraphics[scale=0.65]{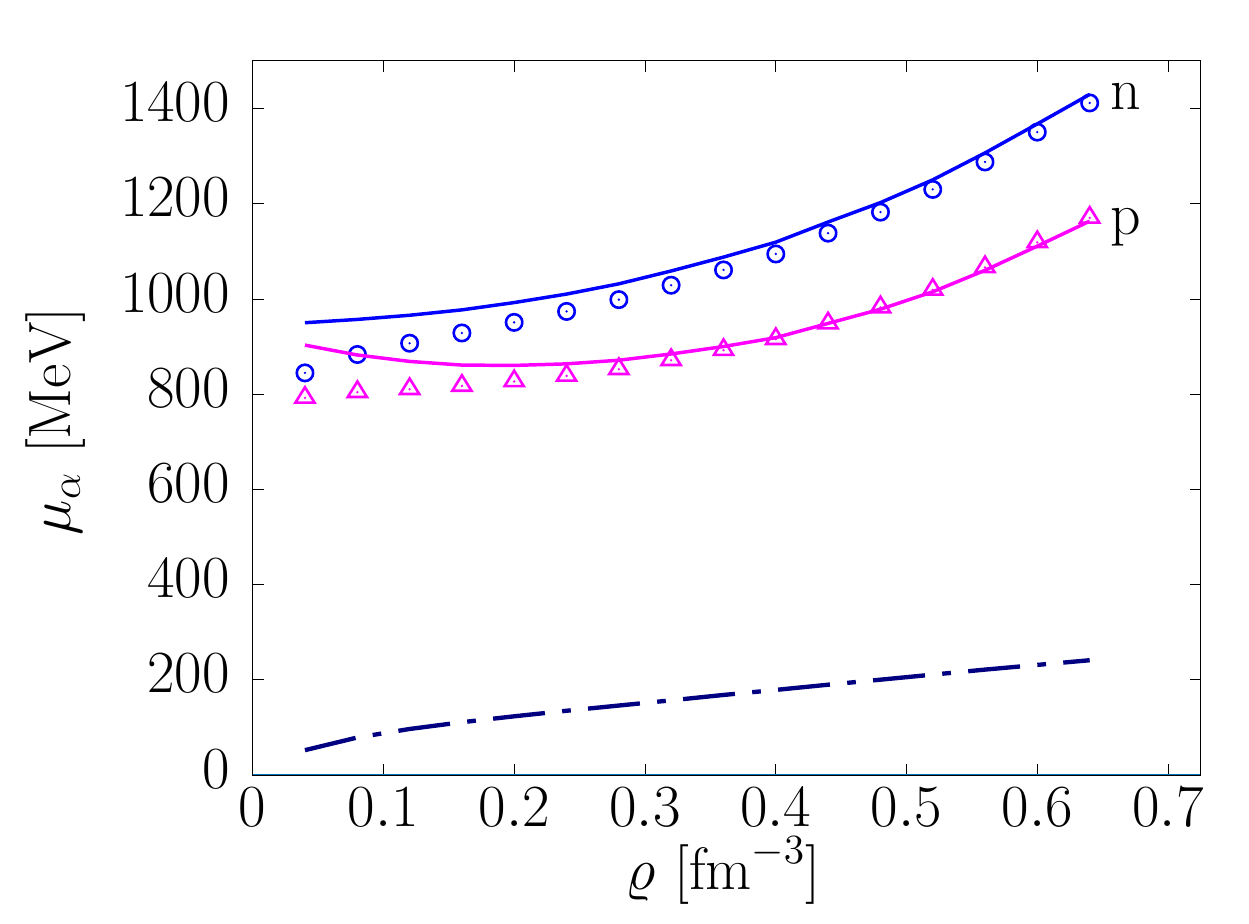} 
\caption{Density dependence of the chemical potentials of protons (p) and neutrons (n) in $\beta$-stable matter 
ar $T=$~50~MeV. For comparison, the corresponding quantities at $T=0$ are shown by the solid lines.
The dot-dash line represents the difference $\mu_n - \mu_p$ at $T=$ 50~MeV.}
\label{display:rhodep}
\end{figure}

The chemical potentials of protons and neutrons in charge-neutral $\beta$-stable matter at 
temperature $T$ = 0 and 50 MeV are displayed in Fig.~\ref{display:rhodep} as a function of baryon density.
For comparison, the difference $\mu_n - \mu_p = \mu_e$ is also shown. 

%%%%%%%%%%%%%%%%%%%%
%%%%%%%%%%%%%%%%%%%%

\begin{figure}[h]
%\vspace{0.5cm}
\includegraphics[scale=0.65]{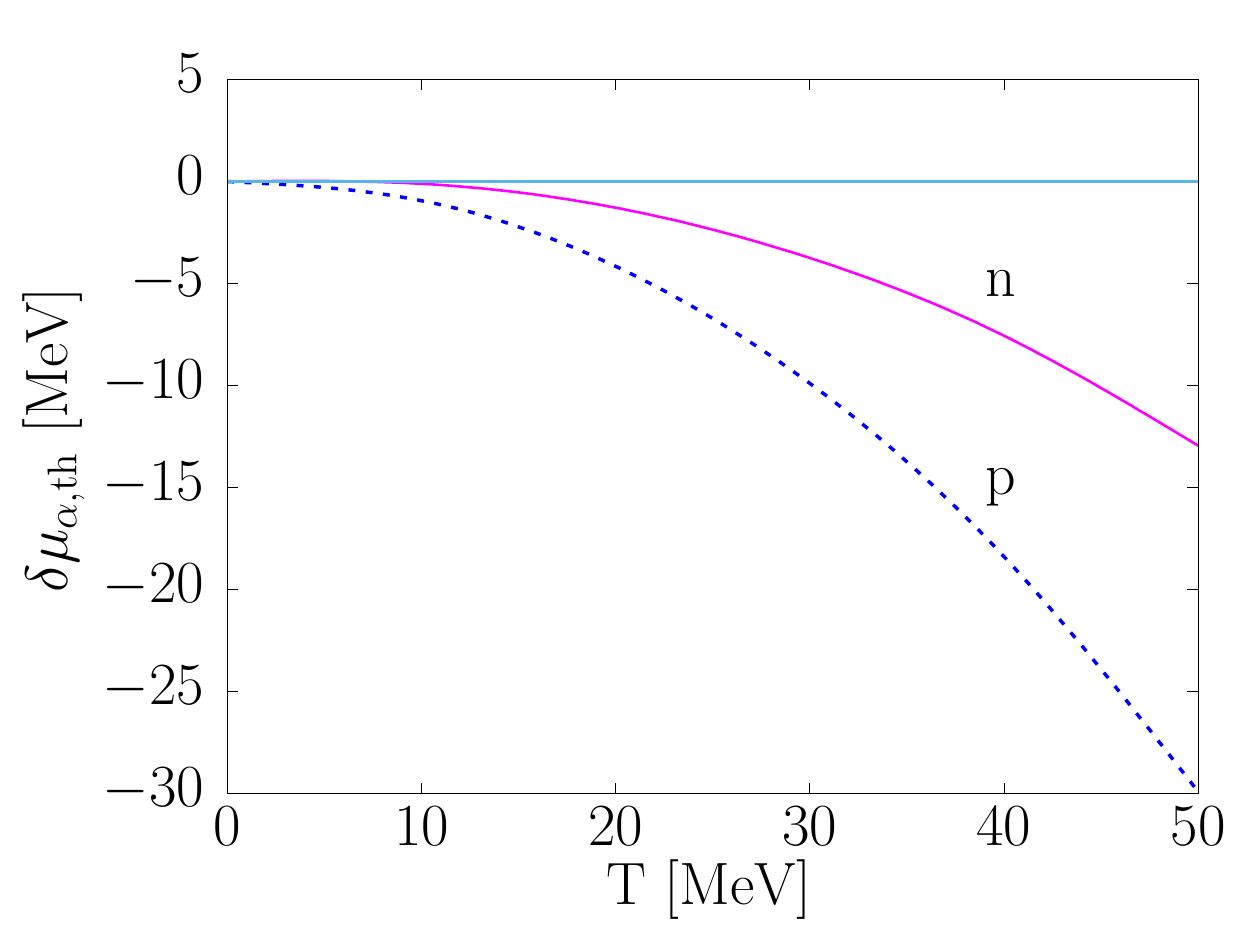} 
\caption{Temperature dependence of  the therrmal contribution to the  proton (p) and neutron (n) chemical potentials, defined by Eq.~\eqref{th:chempot}, in $\beta$-stable matter at baryon density $\varrho = 2 \varrho_0$.}
\label{display:rhoped}
\label{thermal:chempot}
\end{figure}

Thermal effects on chemical potentials can be analysed considering the difference 
\begin{align}
\label{th:chempot}
\delta \mu_{\alpha, {\rm th}} = \mu_\alpha - \mu_{\alpha, 0}\ , 
\end{align}
with $ \mu_{\alpha, 0}$ being the value of  $\mu_\alpha$ in cold matter at fixed baryon density $\varrho$ and particle fraction $Y_\alpha$. Figure~\ref{thermal:chempot} illustrates the temperature dependence of $\delta \mu_{n, {\rm th}}$ in charge-neutral $\beta$-stable matter at baryon density $\varrho = 2 \varrho_0$.

Because thermal effects in $\beta$-stable matter have different impact on proton and neutron properties, the capability  
to accurately predict $\beta$-equilibrium and matter composition using FCA must be carefully investigated. 
The results of numerical calculations carried out within our approach indicate for temperatures up to $T=50$~MeV the discrepancy 
between the proton fractions obtained from FCA and the exact results never exceeds $\sim 3$\% over the considered
range of baryon density.

%%%%%%%%%%%%%%%%%%%%
%%%%%%%%%%%%%%%%%%%%

\subsection{Internal energy and free energy}

\begin{figure}[h]
\includegraphics[scale=0.7]{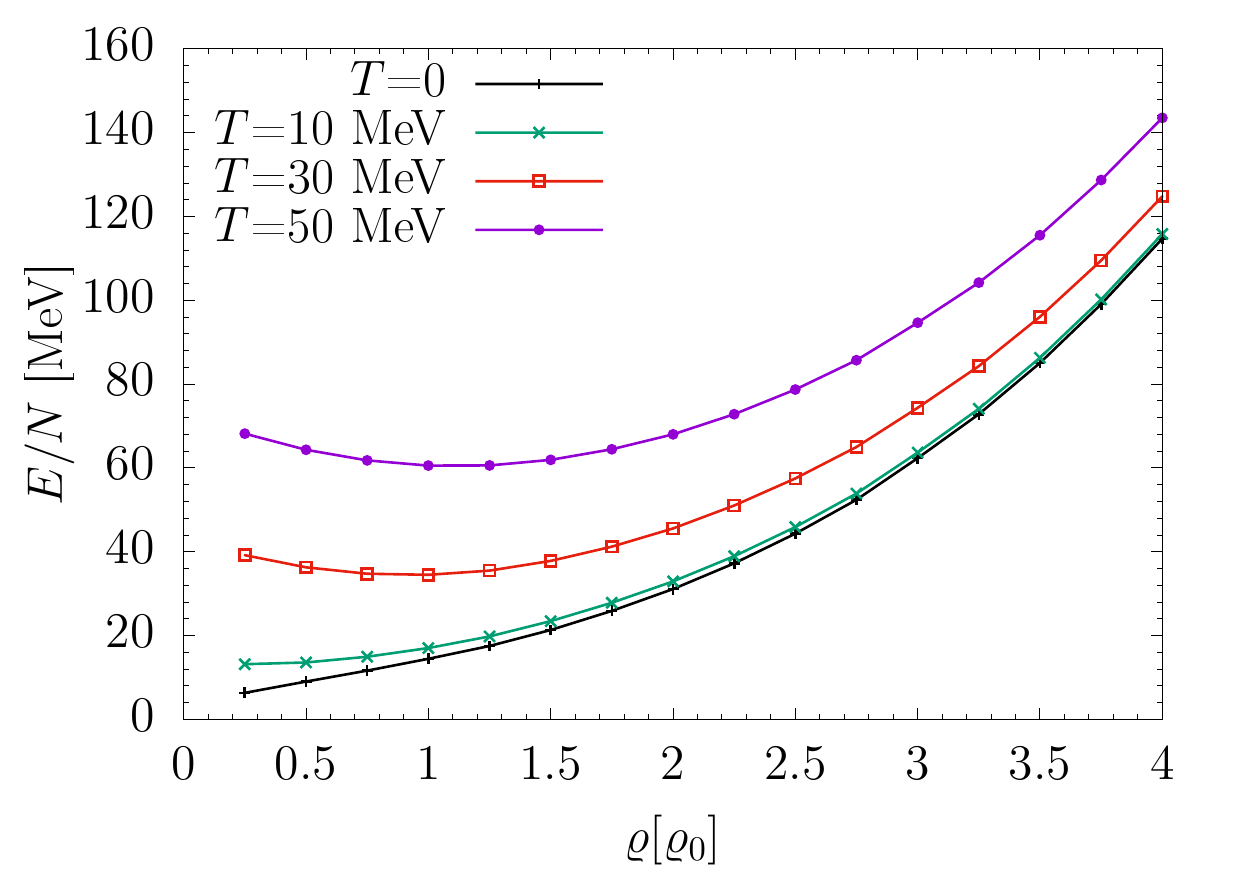}
%vspace*{-.1in}
\caption{Internal energy perr baryon of beta-stable matter as a function of baryon density for different temperatures.
Note that $\varrho$ is given in units of the equilibrium density of cold isospin-symmetric matter, 
$\varrho_0 = 0.16 \ {\rm fm}^{-3}$.}
\label{fig:intenergy_beta}
\end{figure}

\begin{figure}[h]
\includegraphics[scale=0.7]{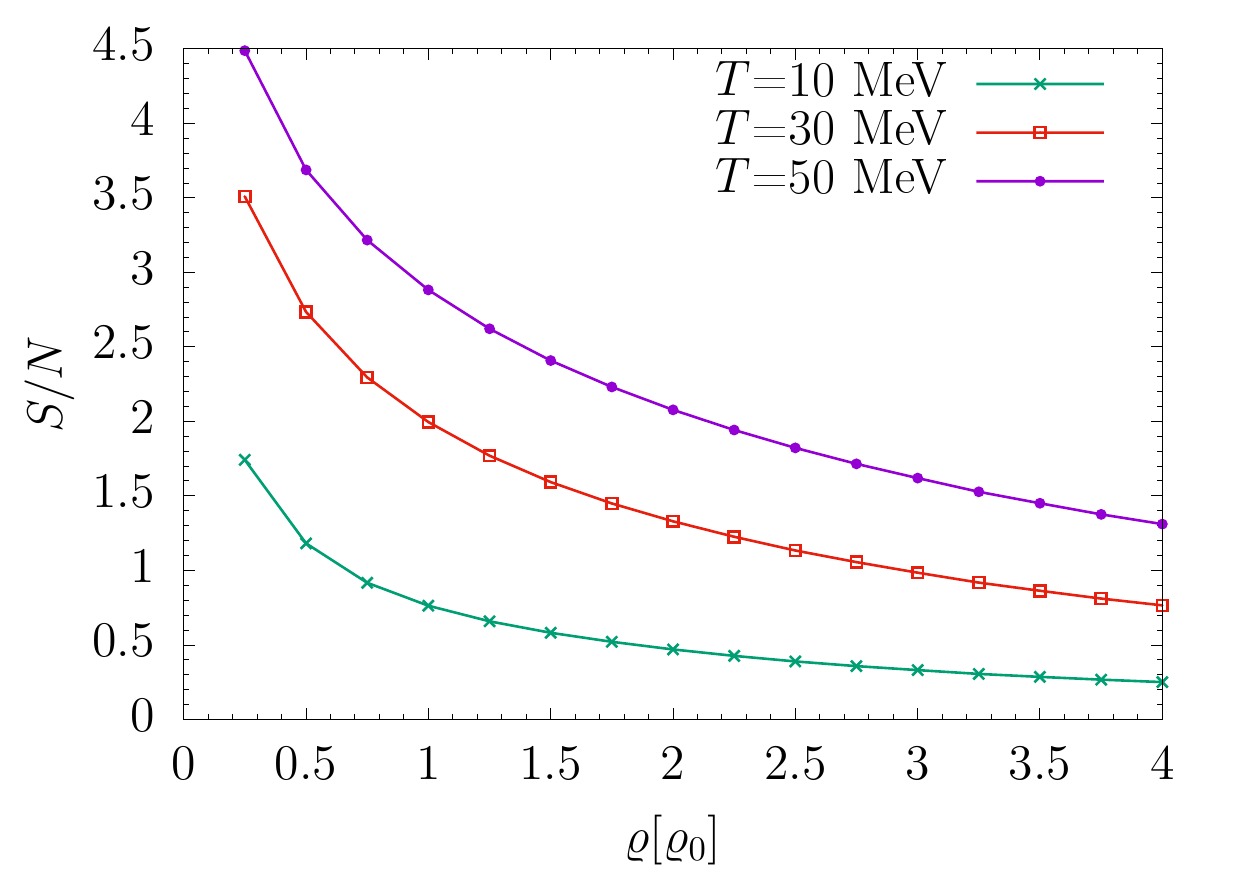}
%vspace*{-.1in}
\caption{Entropy per baryon of beta-stable matter as a function of density for different temperatures.
Note that $\varrho$ is given in units of the equilibrium density of cold isospin-symmetric matter, 
$\varrho_0 = 0.16 \ {\rm fm}^{-3}$.}
\label{fig:entropy_beta}
\end{figure}

The density and temperature dependence of the internal energy and entropy per baryon of $\beta$-stable matter, 
defined according to  Eqs.~\eqref{int:en} and~\eqref{def:S}, respectively, is illustrated in Figs.~\ref{fig:intenergy_beta} and \ref{fig:entropy_beta}. 
It is worth reminding that the CBF effective interaction based on the AV6P+UIX nuclear Hamiltonian yields a remarkably accurate 
account of the equilibrium properties of isospin symmetric matter at $T=0$ inferred from nuclear data. 
The saturation density is correctly reproduced, and the contribution of interactions to the internal energy turns 
out to be within $\sim 10$\% of the empirical value.

%%%%%%%%%%%%%%%%%%%%
%%%%%%%%%%%%%%%%%%%%

Figure~\ref{fig:intenergy_beta} shows that, for any given $\varrho$, the internal energy is an increasing function 
of temperature. However, the concurrent increment of  the proton fraction with $T$, discussed in Section~\ref{composition}, leads 
to the appearance of a minimum for temperatures larger than 10 MeV. 

As expected, thermal contributions to the internal energy turn out to be less important at higher $\varrho$. 
However, for $T>10$~MeV they are still significant at  densities as high as $4 \varrho_0$.
 
%%%%%%%%%%%%%%%%%%%%
%%%%%%%%%%%%%%%%%%%%

%----------------------------------------
\section{Modelling thermal effects} \label{sec:parametrisation}
%----------------------------------------

The description of thermal effects on the thermodynamic functions determining the EOS, that is, pressure and energy density, 
is of paramount importance in view of astrophysical applications. The number of available EOSs of nuclear matter at $T \neq 0$  
is much smaller when compared to the corresponding figure for cold matter. Moreover, the implementation of microscopic 
EOSs in numerical simulation of processes such as binary neutron star merger involves non trivial difficulties. 

This above problems are 
often circumvented using simple but physically sound parametrisation of the EOSs.
An extensively used expression is based on the so-called ''hybrid-EOS'' approach, 
in which thermal contributions to pressure and energy density are described using an approximation based on the 
the ideal fluid law; see Eq.~\eqref{hybrid}. 

As pointed out in the previous section, the results of microscopic calculations clearly signal a strong interplay between the 
dependencies of the nuclear matter properties on density and temperature. This feature obviously questions the adequacy of the assumption that thermal contributions to the EOS be the same to all densities. Motivated by this observation, Raithel {\it et al.} have recently proposed a model that explicitly takes into consideration the effect of matter degeneracy~\cite{raithel2019}. 

Rather than using the ideal fluid EOS
in the whole density range, the authors of Ref.~\cite{raithel2019} employ the Sommerfeld expansion described by  \citet{constantinou2015} in the region high $\varrho$. This formalism allows to write the deviations of the thermodynamic functions 
from their zero-temperature values as series of powers of  $T$. The calculation of the next-to-leading order term 
involves the nucleon effective mass and its derivatives, which implies that a model of nuclear dynamics at $T\neq0$  is needed beforehand. 

In order to make their parametrisation as general as possible, \citet{raithel2019} considered a set of RMF models for which the effective masses at different temperatures are available in the literature, and performed a fit using analytical models, such as 
piecewise polytropes, as zero-temperature baseline.

Our goal here is establish the extent to which the results reported in Ref.~\cite{raithel2019} stand, when compared to an EOS
obtained within the framework of NMBT, rather than the RMF approach. We use the parameter values $n_0 \sim 0.13 \ \mathrm{fm}^{-3}$ and 
$\alpha \sim 0.9$ \textemdash see Box 1 of Ref.~\citep{raithel2019} and the erratum, Ref.~\citep{raithel2019erratum}\textemdash to obtain first the effective mass, and subsequently the internal energy per baryon and the matter pressure. 
Note that the results reported in Ref.~\cite{raithel2019} do not include the contribution of muons. Therefore, our analysis will be 
limited to the case of $npe$ matter.

\begin{figure}[!h]
\includegraphics[scale=0.7]{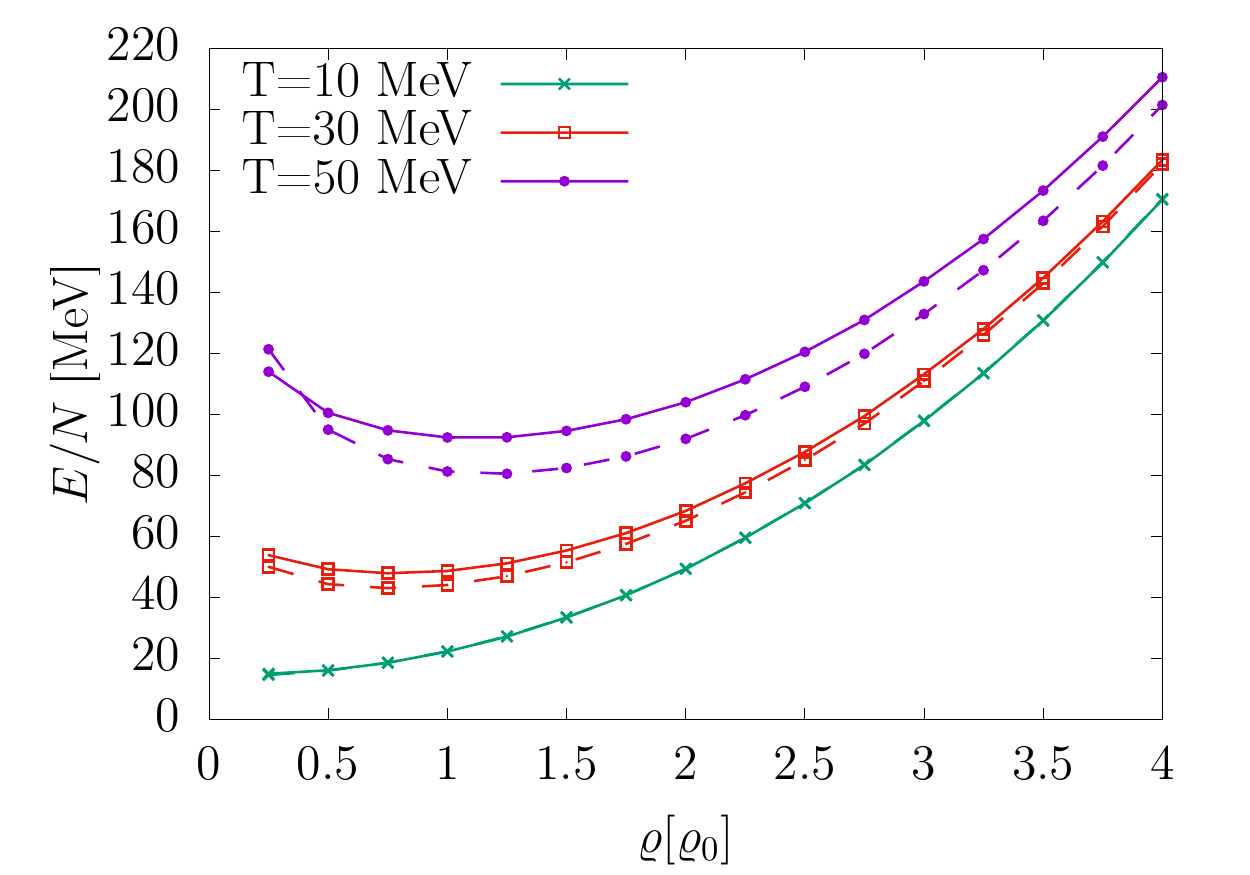}
%vspace*{-.1in}
\caption{Density dependence of the internal energy per baryon of $npe$ matter in $\beta$-equilibrium. Solid and dashed lines  
correspond to the results of our calculations and to the fit of Refs.~\citep{raithel2019,raithel2019erratum}, }
\label{fig:E_raithel}
\end{figure}

In Fig. \ref{fig:E_raithel} we show a comparison between the internal energy per baryon of $\beta$-stable $npe$ matter 
obtained from the approach described in the previous section (solid lines) and the fit of Refs.~\citep{raithel2019,raithel2019erratum} (dashed lines). It is apparent that at $T = \mathrm{10 \ MeV}$, the agreement is almost perfect, while discrepancies\textemdash the size of which increases with increasing $T$\textemdash are clearly visible at larger temperatures. The maximum relative error between the fit and the miscroscopic calculation at $T = \mathrm{50 \ MeV}$ ($\mathrm{30 \ MeV}$) turns out to be  $\sim 16\%$ ($\sim 11\%$),  and occurs at density $\sim 1.5 \varrho_0$ ($\sim \varrho_0$).

\begin{figure}[!h]
\includegraphics[scale=0.7]{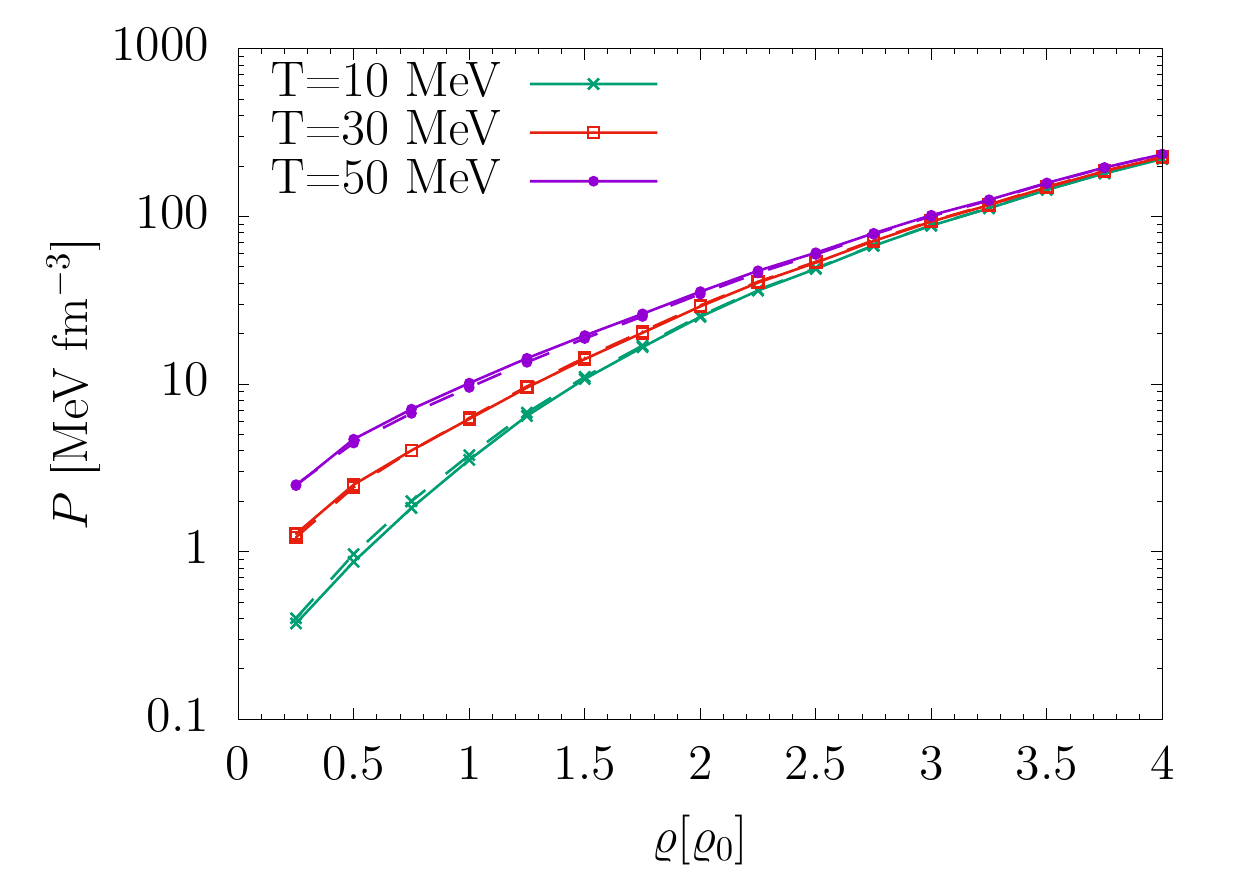}
%vspace*{-.1in}
\caption{Comparison between the pressure of $\beta$-stable $npe$ matter obtained using the approximate model of Ref.~\citep{raithel2019,raithel2019erratum} (dashed lines) and the microscopic approach described in this 
work (solid lines).}
\label{fig:P_raithel}
\end{figure}

We have also analysed the accuracy of the approximation of \citet{raithel2019} for the pressure. 
A comparison with the results obtained from our microscopic approach, illustrated in  Fig.~\ref{fig:P_raithel}, shows
a remarkably good agreement over the whole temperature range. The parametrisation of Ref.\cite{raithel2019} appear to 
properly take into account the effects of degeneracy at all densities.

\begin{figure}[!th]
\includegraphics[scale=0.7]{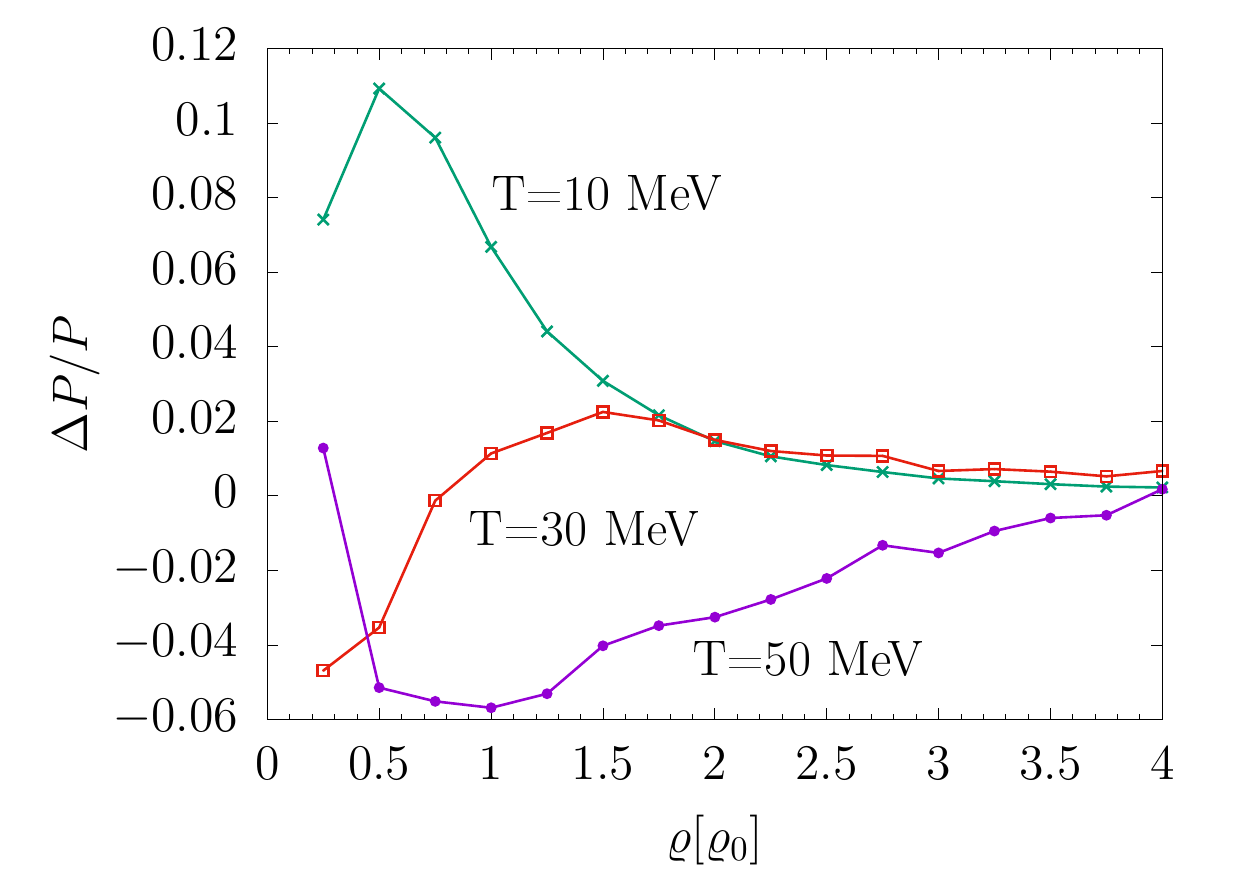}
%vspace*{-.1in}
\caption{Relative difference between the pressure of $\beta$-stable $npe$ matter obtained 
using the model of \citet{raithel2021} and that resulting from the microscopic approach described in this work.}
\label{fig:diffP_raithel}
\end{figure}

In order to provide a quantitative estimate of the validity of the the approximations involved in the parametrisation of pressure, 
in Fig.~\ref{fig:diffP_raithel} we report the relative difference
\begin{align}
\label{diffP}
\frac{\Delta P}{P} = \frac{(P_\mathrm{approx} - P)}{P} \ , 
\end{align}
where $P$ is the result of our calculation, as a function of baryon density. It is apparent that the largest errors occur at low densities, and never exceeds $\sim$6\% for $\varrho > \varrho_0$.

%It is apparent that for $T = 10 \ \mathrm{MeV}$ the error is always less than $\sim$3\%, and never exceeds $\sim$13\% for temperatures up to 50 MeV.

%%%%%%%%%%%%%%%%%%%%
%%%%%%%%%%%%%%%%%%%%

%%%%%%%%%%%%%%%%%%%%
%%%%%%%%%%%%%%%%%%%%

%%%%%%%%%%%%%%%%%%%%
%\begin{figure}[!h]
%%%%%%%%%%%%%%%%%%%%

%----------------------------------------
\section{Summary and Conclusions} \label{sec:conclusion}
%----------------------------------------

We have analysed the impact of temperature on several properties of charge-neutral 
nuclear matter in $\beta$-equilibrium. Calculations have been performed using
the formalism of finite-temperature perturbation theory,  with an effective interaction 
derived from a nuclear Hamiltonian comprising both two- and three-nucleon potentials.

The most prominent feature emerging from our results is the strong interplay between 
temperature and density, that can be ultimately traced back to the form of the Fermi 
distribution. For any given temperature thermal effects turn out to decrease with density, 
although in some instances they are still significant at density as high as $\sim 4\varrho$.
As a consequence, in $\beta$-stable matter thermal modifications of nucleon properties, 
such as the energy spectrum, are more pronounced for protons than for neutrons.  

The interplay of temperature and density has no trivial implications for astrophysical studies. 
The temperature and density profiles obtained from neutron star merger simulations\textemdash see, e.g. Refs.~\citep{figura2020,camelio2021,raithel2021}), showing that in the inner region of the remnant 
the thermal contribution to the pressure is lower. However, this happens not only because the 
degeneracy pressure becomes more important, but also because the temperature is lower. 
On the other hand, at intermediate densities the temperature is higher and the thermal 
contribution to the pressure is larger as well. At lower densities, despite the temperature being 
lower, the thermal contribution to the pressure is even more important due to naturally lower 
degeneracy pressure. It clearly appears that, in order to pin down the role of thermal effects in determining 
the properties of neutron star matter, their temperature and density dependence must be accurately 
described within a consistent framework.

Of great importance, in this context, will be the availability of simple parametrisations of the EOS 
of hot nuclear matter in $\beta$-equilibrium, suitable for use in numerical simulations. A direct comparison 
to the results of our calculations shows that the approximate treatment of thermal effects recently proposed by  
\citet{raithel2021} is remarkably accurate, and suitable to describe EOSs obtained from different models
of nuclear dynamics.

It is important to keep in mind that the discussion of temperature effects in nuclear matter should not be 
limited to thermal contributions to average properties, such as the pressure and energy density. As shown by the results discussed in this article, the most fundamental properties, including the Fermi distributions, single-particle spectra and effective masses, are significantly modified at finite temperature. A consistent inclusion of the temperature dependence of these quantities is essential to accurately describe nuclear collision rates in matter, which in turn determine out-of-equilibrium phenomena \citep{alford2018,alford2021,mostetal2021,mostetal2022}, as well as neutrino emission. 
The approach described in this article allows to carry out calculations of, e.g., the rates of modified Urca processes at $T>0$, 
using nuclear matrix elements obtained from a highly realistic nuclear Hamiltonian, comprising both two- and three-nucleon 
potentials. 

As a final remark, is should be also mentioned that in this work we have considered $\beta$-equilibrium in the absence of 
neutrinos. However, neutrino trapping is expected to occur even at $T \lesssim \mathrm{10 \ MeV}$ \citep{alford2018beta}, and 
we plan to extend our calculations to study this scenario.

%%%%%%%%%%%%%%%%%%%%%%%%%%%%%%%%%%%%%%%%%%%%%%%%%%%%%
\acknowledgments
This work has been supported by the Italian National Institute for Nuclear
Research (INFN) under grant TEONGRAV. 
%OB gratefully acknowledges the kind hospitality of the 
%Center for Neutrino Physics at Virginia Tech, where part of this work has been carried out. 
%
%\bibliographystyle{apsrev4-1}
%\bibliography{paper}

\input{Tonetto_Benhar.bbl}

\end{document}

%% file: Tonetto_Benhar.bbl

%% file: Tonetto_Benhar.bbl
\begin{thebibliography}{43}%
\makeatletter
\providecommand \@ifxundefined [1]{%
 \@ifx{#1\undefined}
}%
\providecommand \@ifnum [1]{%
 \ifnum #1\expandafter \@firstoftwo
 \else \expandafter \@secondoftwo
 \fi
}%
\providecommand \@ifx [1]{%
 \ifx #1\expandafter \@firstoftwo
 \else \expandafter \@secondoftwo
 \fi
}%
\providecommand \natexlab [1]{#1}%
\providecommand \enquote  [1]{``#1''}%
\providecommand \bibnamefont  [1]{#1}%
\providecommand \bibfnamefont [1]{#1}%
\providecommand \citenamefont [1]{#1}%
\providecommand \href@noop [0]{\@secondoftwo}%
\providecommand \href [0]{\begingroup \@sanitize@url \@href}%
\providecommand \@href[1]{\@@startlink{#1}\@@href}%
\providecommand \@@href[1]{\endgroup#1\@@endlink}%
\providecommand \@sanitize@url [0]{\catcode `\\12\catcode `\$12\catcode
  `\&12\catcode `\#12\catcode `\^12\catcode `\_12\catcode `\%12\relax}%
\providecommand \@@startlink[1]{}%
\providecommand \@@endlink[0]{}%
\providecommand \url  [0]{\begingroup\@sanitize@url \@url }%
\providecommand \@url [1]{\endgroup\@href {#1}{\urlprefix }}%
\providecommand \urlprefix  [0]{URL }%
\providecommand \Eprint [0]{\href }%
\providecommand \doibase [0]{http://dx.doi.org/}%
\providecommand \selectlanguage [0]{\@gobble}%
\providecommand \bibinfo  [0]{\@secondoftwo}%
\providecommand \bibfield  [0]{\@secondoftwo}%
\providecommand \translation [1]{[#1]}%
\providecommand \BibitemOpen [0]{}%
\providecommand \bibitemStop [0]{}%
\providecommand \bibitemNoStop [0]{.\EOS\space}%
\providecommand \EOS [0]{\spacefactor3000\relax}%
\providecommand \BibitemShut  [1]{\csname bibitem#1\endcsname}%
\let\auto@bib@innerbib\@empty
%</preamble>
\bibitem [{\citenamefont {{Burrows}}\ and\ \citenamefont
  {{Lattimer}}(1986)}]{burrows1986}%
  \BibitemOpen
  \bibfield  {author} {\bibinfo {author} {\bibfnamefont {A.}~\bibnamefont
  {{Burrows}}}\ and\ \bibinfo {author} {\bibfnamefont {J.~M.}\ \bibnamefont
  {{Lattimer}}},\ }\href {\doibase 10.1086/164405} {\bibfield  {journal}
  {\bibinfo  {journal} {The Astrophysical Journal}\ }\textbf {\bibinfo {volume}
  {307}},\ \bibinfo {pages} {178} (\bibinfo {year} {1986})}\BibitemShut
  {NoStop}%
\bibitem [{\citenamefont {Keil}\ and\ \citenamefont {Janka}(1995)}]{keil1995}%
  \BibitemOpen
  \bibfield  {author} {\bibinfo {author} {\bibfnamefont {W.}~\bibnamefont
  {Keil}}\ and\ \bibinfo {author} {\bibfnamefont {H.~T.}\ \bibnamefont
  {Janka}},\ }\href@noop {} {\bibfield  {journal} {\bibinfo  {journal} {\aap}\
  }\textbf {\bibinfo {volume} {296}},\ \bibinfo {pages} {145} (\bibinfo {year}
  {1995})}\BibitemShut {NoStop}%
\bibitem [{\citenamefont {Pons}\ \emph {et~al.}(1999)\citenamefont {Pons},
  \citenamefont {Reddy}, \citenamefont {Prakash}, \citenamefont {Lattimer},\
  and\ \citenamefont {Miralles}}]{pons1999}%
  \BibitemOpen
  \bibfield  {author} {\bibinfo {author} {\bibfnamefont {J.~A.}\ \bibnamefont
  {Pons}}, \bibinfo {author} {\bibfnamefont {S.}~\bibnamefont {Reddy}},
  \bibinfo {author} {\bibfnamefont {M.}~\bibnamefont {Prakash}}, \bibinfo
  {author} {\bibfnamefont {J.~M.}\ \bibnamefont {Lattimer}}, \ and\ \bibinfo
  {author} {\bibfnamefont {J.~A.}\ \bibnamefont {Miralles}},\ }\href {\doibase
  10.1086/306889} {\bibfield  {journal} {\bibinfo  {journal} {The Astrophysical
  Journal}\ }\textbf {\bibinfo {volume} {513}},\ \bibinfo {pages} {780}
  (\bibinfo {year} {1999})}\BibitemShut {NoStop}%
\bibitem [{\citenamefont {Camelio}\ \emph {et~al.}(2017)\citenamefont
  {Camelio}, \citenamefont {Lovato}, \citenamefont {Gualtieri}, \citenamefont
  {Benhar}, \citenamefont {Pons},\ and\ \citenamefont {Ferrari}}]{camelio2017}%
  \BibitemOpen
  \bibfield  {author} {\bibinfo {author} {\bibfnamefont {G.}~\bibnamefont
  {Camelio}}, \bibinfo {author} {\bibfnamefont {A.}~\bibnamefont {Lovato}},
  \bibinfo {author} {\bibfnamefont {L.}~\bibnamefont {Gualtieri}}, \bibinfo
  {author} {\bibfnamefont {O.}~\bibnamefont {Benhar}}, \bibinfo {author}
  {\bibfnamefont {J.~A.}\ \bibnamefont {Pons}}, \ and\ \bibinfo {author}
  {\bibfnamefont {V.}~\bibnamefont {Ferrari}},\ }\href {\doibase
  10.1103/PhysRevD.96.043015} {\bibfield  {journal} {\bibinfo  {journal} {Phys.
  Rev. D}\ }\textbf {\bibinfo {volume} {96}},\ \bibinfo {pages} {043015}
  (\bibinfo {year} {2017})}\BibitemShut {NoStop}%
\bibitem [{\citenamefont {Baiotti}\ and\ \citenamefont
  {Rezzolla}(2017)}]{baiotti2017}%
  \BibitemOpen
  \bibfield  {author} {\bibinfo {author} {\bibfnamefont {L.}~\bibnamefont
  {Baiotti}}\ and\ \bibinfo {author} {\bibfnamefont {L.}~\bibnamefont
  {Rezzolla}},\ }\href {\doibase 10.1088/1361-6633/aa67bb} {\bibfield
  {journal} {\bibinfo  {journal} {Rep. Prog. in Phys.}\ }\textbf {\bibinfo
  {volume} {80}},\ \bibinfo {pages} {096901} (\bibinfo {year}
  {2017})}\BibitemShut {NoStop}%
\bibitem [{\citenamefont {Raithel}\ \emph
  {et~al.}(2021{\natexlab{a}})\citenamefont {Raithel}, \citenamefont
  {Paschalidis},\ and\ \citenamefont {\"Ozel}}]{raithel2021}%
  \BibitemOpen
  \bibfield  {author} {\bibinfo {author} {\bibfnamefont {C.~A.}\ \bibnamefont
  {Raithel}}, \bibinfo {author} {\bibfnamefont {V.}~\bibnamefont
  {Paschalidis}}, \ and\ \bibinfo {author} {\bibfnamefont {F.}~\bibnamefont
  {\"Ozel}},\ }\href {\doibase 10.1103/PhysRevD.104.063016} {\bibfield
  {journal} {\bibinfo  {journal} {Phys. Rev. D}\ }\textbf {\bibinfo {volume}
  {104}},\ \bibinfo {pages} {063016} (\bibinfo {year}
  {2021}{\natexlab{a}})}\BibitemShut {NoStop}%
\bibitem [{\citenamefont {Figura}\ \emph {et~al.}(2020)\citenamefont {Figura},
  \citenamefont {Lu}, \citenamefont {Burgio}, \citenamefont {Li},\ and\
  \citenamefont {Schulze}}]{figura2020}%
  \BibitemOpen
  \bibfield  {author} {\bibinfo {author} {\bibfnamefont {A.}~\bibnamefont
  {Figura}}, \bibinfo {author} {\bibfnamefont {J.-J.}\ \bibnamefont {Lu}},
  \bibinfo {author} {\bibfnamefont {G.~F.}\ \bibnamefont {Burgio}}, \bibinfo
  {author} {\bibfnamefont {Z.-H.}\ \bibnamefont {Li}}, \ and\ \bibinfo {author}
  {\bibfnamefont {H.-J.}\ \bibnamefont {Schulze}},\ }\href {\doibase
  10.1103/PhysRevD.102.043006} {\bibfield  {journal} {\bibinfo  {journal}
  {Phys. Rev. D}\ }\textbf {\bibinfo {volume} {102}},\ \bibinfo {pages}
  {043006} (\bibinfo {year} {2020})}\BibitemShut {NoStop}%
\bibitem [{\citenamefont {Figura}\ \emph {et~al.}(2021)\citenamefont {Figura},
  \citenamefont {Li}, \citenamefont {Lu}, \citenamefont {Burgio}, \citenamefont
  {Li},\ and\ \citenamefont {Schulze}}]{figura2021}%
  \BibitemOpen
  \bibfield  {author} {\bibinfo {author} {\bibfnamefont {A.}~\bibnamefont
  {Figura}}, \bibinfo {author} {\bibfnamefont {F.}~\bibnamefont {Li}}, \bibinfo
  {author} {\bibfnamefont {J.-J.}\ \bibnamefont {Lu}}, \bibinfo {author}
  {\bibfnamefont {G.~F.}\ \bibnamefont {Burgio}}, \bibinfo {author}
  {\bibfnamefont {Z.-H.}\ \bibnamefont {Li}}, \ and\ \bibinfo {author}
  {\bibfnamefont {H.-J.}\ \bibnamefont {Schulze}},\ }\href {\doibase
  10.1103/PhysRevD.103.083012} {\bibfield  {journal} {\bibinfo  {journal}
  {Phys. Rev. D}\ }\textbf {\bibinfo {volume} {103}},\ \bibinfo {pages}
  {083012} (\bibinfo {year} {2021})}\BibitemShut {NoStop}%
\bibitem [{\citenamefont {Hammond}\ \emph {et~al.}(2021)\citenamefont
  {Hammond}, \citenamefont {Hawke},\ and\ \citenamefont
  {Andersson}}]{hammond2021}%
  \BibitemOpen
  \bibfield  {author} {\bibinfo {author} {\bibfnamefont {P.}~\bibnamefont
  {Hammond}}, \bibinfo {author} {\bibfnamefont {I.}~\bibnamefont {Hawke}}, \
  and\ \bibinfo {author} {\bibfnamefont {N.}~\bibnamefont {Andersson}},\ }\href
  {\doibase 10.1103/PhysRevD.104.103006} {\bibfield  {journal} {\bibinfo
  {journal} {Phys. Rev. D}\ }\textbf {\bibinfo {volume} {104}},\ \bibinfo
  {pages} {103006} (\bibinfo {year} {2021})}\BibitemShut {NoStop}%
\bibitem [{\citenamefont {Alford}\ \emph {et~al.}(2018)\citenamefont {Alford},
  \citenamefont {Bovard}, \citenamefont {Hanauske}, \citenamefont {Rezzolla},\
  and\ \citenamefont {Schwenzer}}]{alford2018}%
  \BibitemOpen
  \bibfield  {author} {\bibinfo {author} {\bibfnamefont {M.~G.}\ \bibnamefont
  {Alford}}, \bibinfo {author} {\bibfnamefont {L.}~\bibnamefont {Bovard}},
  \bibinfo {author} {\bibfnamefont {M.}~\bibnamefont {Hanauske}}, \bibinfo
  {author} {\bibfnamefont {L.}~\bibnamefont {Rezzolla}}, \ and\ \bibinfo
  {author} {\bibfnamefont {K.}~\bibnamefont {Schwenzer}},\ }\href {\doibase
  10.1103/PhysRevLett.120.041101} {\bibfield  {journal} {\bibinfo  {journal}
  {Phys. Rev. Lett.}\ }\textbf {\bibinfo {volume} {120}},\ \bibinfo {pages}
  {041101} (\bibinfo {year} {2018})}\BibitemShut {NoStop}%
\bibitem [{Com()}]{CompOSE}%
  \BibitemOpen
  \href@noop {} {\enquote {\bibinfo {title} {{CompOSE: CompStar Online
  Supernov\ae \ Equations of State}},}\ }\bibinfo {howpublished}
  {\url{https://compose.obspm.fr/articles}}\BibitemShut {NoStop}%
\bibitem [{\citenamefont {Raithel}\ \emph {et~al.}(2019)\citenamefont
  {Raithel}, \citenamefont {\"Ozel},\ and\ \citenamefont
  {Paschalidis}}]{raithel2019}%
  \BibitemOpen
  \bibfield  {author} {\bibinfo {author} {\bibfnamefont {C.~A.}\ \bibnamefont
  {Raithel}}, \bibinfo {author} {\bibfnamefont {F.}~\bibnamefont {\"Ozel}}, \
  and\ \bibinfo {author} {\bibfnamefont {V.}~\bibnamefont {Paschalidis}},\
  }\href {\doibase 10.3847/1538-4357/ab08ea} {\bibfield  {journal} {\bibinfo
  {journal} {The Astrophysical Journal}\ }\textbf {\bibinfo {volume} {875}},\
  \bibinfo {pages} {12} (\bibinfo {year} {2019})}\BibitemShut {NoStop}%
\bibitem [{\citenamefont {Oechslin}\ \emph {et~al.}(2007)\citenamefont
  {Oechslin}, \citenamefont {Janka},\ and\ \citenamefont
  {Marek}}]{oechslin2007}%
  \BibitemOpen
  \bibfield  {author} {\bibinfo {author} {\bibfnamefont {R.}~\bibnamefont
  {Oechslin}}, \bibinfo {author} {\bibfnamefont {H.-T.}\ \bibnamefont {Janka}},
  \ and\ \bibinfo {author} {\bibfnamefont {A.}~\bibnamefont {Marek}},\ }\href
  {\doibase 10.1051/0004-6361:20066682} {\bibfield  {journal} {\bibinfo
  {journal} {A\&A}\ }\textbf {\bibinfo {volume} {467}},\ \bibinfo {pages} {395}
  (\bibinfo {year} {2007})}\BibitemShut {NoStop}%
\bibitem [{\citenamefont {Prakash}\ \emph {et~al.}(1997)\citenamefont
  {Prakash}, \citenamefont {Bombaci}, \citenamefont {Prakash}, \citenamefont
  {Ellis}, \citenamefont {Lattimer},\ and\ \citenamefont
  {Knorren}}]{Prakash:1997}%
  \BibitemOpen
  \bibfield  {author} {\bibinfo {author} {\bibfnamefont {M.}~\bibnamefont
  {Prakash}}, \bibinfo {author} {\bibfnamefont {I.}~\bibnamefont {Bombaci}},
  \bibinfo {author} {\bibfnamefont {M.}~\bibnamefont {Prakash}}, \bibinfo
  {author} {\bibfnamefont {P.~J.}\ \bibnamefont {Ellis}}, \bibinfo {author}
  {\bibfnamefont {J.~M.}\ \bibnamefont {Lattimer}}, \ and\ \bibinfo {author}
  {\bibfnamefont {R.}~\bibnamefont {Knorren}},\ }\href@noop {} {\bibfield
  {journal} {\bibinfo  {journal} {Phys. Rep.}\ }\textbf {\bibinfo {volume}
  {280}},\ \bibinfo {pages} {1} (\bibinfo {year} {1997})}\BibitemShut {NoStop}%
\bibitem [{\citenamefont {Kaplan}\ \emph {et~al.}(2014)\citenamefont {Kaplan},
  \citenamefont {Ott}, \citenamefont {O'Connor}, \citenamefont {Kiuchi},
  \citenamefont {Roberts},\ and\ \citenamefont {Duez}}]{Kaplan:2014}%
  \BibitemOpen
  \bibfield  {author} {\bibinfo {author} {\bibfnamefont {J.~D.}\ \bibnamefont
  {Kaplan}}, \bibinfo {author} {\bibfnamefont {C.~D.}\ \bibnamefont {Ott}},
  \bibinfo {author} {\bibfnamefont {E.~P.}\ \bibnamefont {O'Connor}}, \bibinfo
  {author} {\bibfnamefont {K.}~\bibnamefont {Kiuchi}}, \bibinfo {author}
  {\bibfnamefont {L.}~\bibnamefont {Roberts}}, \ and\ \bibinfo {author}
  {\bibfnamefont {M.}~\bibnamefont {Duez}},\ }\href@noop {} {\bibfield
  {journal} {\bibinfo  {journal} {The Astrophysical Journal}\ }\textbf
  {\bibinfo {volume} {790}},\ \bibinfo {pages} {19} (\bibinfo {year}
  {2014})}\BibitemShut {NoStop}%
\bibitem [{\citenamefont {Lu}\ \emph {et~al.}(2019)\citenamefont {Lu},
  \citenamefont {Li}, \citenamefont {Burgio}, \citenamefont {Figura},\ and\
  \citenamefont {Schulze}}]{PhysRevC.100.054335}%
  \BibitemOpen
  \bibfield  {author} {\bibinfo {author} {\bibfnamefont {J.-J.}\ \bibnamefont
  {Lu}}, \bibinfo {author} {\bibfnamefont {Z.-H.}\ \bibnamefont {Li}}, \bibinfo
  {author} {\bibfnamefont {G.~F.}\ \bibnamefont {Burgio}}, \bibinfo {author}
  {\bibfnamefont {A.}~\bibnamefont {Figura}}, \ and\ \bibinfo {author}
  {\bibfnamefont {H.-J.}\ \bibnamefont {Schulze}},\ }\href {\doibase
  10.1103/PhysRevC.100.054335} {\bibfield  {journal} {\bibinfo  {journal}
  {Phys. Rev. C}\ }\textbf {\bibinfo {volume} {100}},\ \bibinfo {pages}
  {054335} (\bibinfo {year} {2019})}\BibitemShut {NoStop}%
\bibitem [{\citenamefont {Benhar}\ \emph {et~al.}()\citenamefont {Benhar},
  \citenamefont {Lovato},\ and\ \citenamefont {Camelio}}]{benhar2022}%
  \BibitemOpen
  \bibfield  {author} {\bibinfo {author} {\bibfnamefont {O.}~\bibnamefont
  {Benhar}}, \bibinfo {author} {\bibfnamefont {A.}~\bibnamefont {Lovato}}, \
  and\ \bibinfo {author} {\bibfnamefont {G.}~\bibnamefont {Camelio}},\
  }\href@noop {} {\enquote {\bibinfo {title} {Modelling neutron star matter in
  the age of multimessenger astrophysics},}\ }\bibinfo {note}
  {"arXiv:2205.XXXXX [nucl-th]"}\BibitemShut {NoStop}%
\bibitem [{\citenamefont {Benhar}\ and\ \citenamefont
  {Lovato}(2017{\natexlab{a}})}]{BL2017}%
  \BibitemOpen
  \bibfield  {author} {\bibinfo {author} {\bibfnamefont {O.}~\bibnamefont
  {Benhar}}\ and\ \bibinfo {author} {\bibfnamefont {A.}~\bibnamefont
  {Lovato}},\ }\href {\doibase 10.1103/PhysRevC.96.054301} {\bibfield
  {journal} {\bibinfo  {journal} {Phys. Rev. C}\ }\textbf {\bibinfo {volume}
  {96}},\ \bibinfo {pages} {054301} (\bibinfo {year}
  {2017}{\natexlab{a}})}\BibitemShut {NoStop}%
\bibitem [{\citenamefont {Bauswein}\ \emph {et~al.}(2010)\citenamefont
  {Bauswein}, \citenamefont {Janka},\ and\ \citenamefont
  {Oechslin}}]{bauswein2010}%
  \BibitemOpen
  \bibfield  {author} {\bibinfo {author} {\bibfnamefont {A.}~\bibnamefont
  {Bauswein}}, \bibinfo {author} {\bibfnamefont {H.-T.}\ \bibnamefont {Janka}},
  \ and\ \bibinfo {author} {\bibfnamefont {R.}~\bibnamefont {Oechslin}},\
  }\href {\doibase 10.1103/PhysRevD.82.084043} {\bibfield  {journal} {\bibinfo
  {journal} {Phys. Rev. D}\ }\textbf {\bibinfo {volume} {82}},\ \bibinfo
  {pages} {084043} (\bibinfo {year} {2010})}\BibitemShut {NoStop}%
\bibitem [{\citenamefont {Hotokezaka}\ \emph {et~al.}(2011)\citenamefont
  {Hotokezaka}, \citenamefont {Kyutoku}, \citenamefont {Okawa}, \citenamefont
  {Shibata},\ and\ \citenamefont {Kiuchi}}]{hotokezaka2011}%
  \BibitemOpen
  \bibfield  {author} {\bibinfo {author} {\bibfnamefont {K.}~\bibnamefont
  {Hotokezaka}}, \bibinfo {author} {\bibfnamefont {K.}~\bibnamefont {Kyutoku}},
  \bibinfo {author} {\bibfnamefont {H.}~\bibnamefont {Okawa}}, \bibinfo
  {author} {\bibfnamefont {M.}~\bibnamefont {Shibata}}, \ and\ \bibinfo
  {author} {\bibfnamefont {K.}~\bibnamefont {Kiuchi}},\ }\href {\doibase
  10.1103/PhysRevD.83.124008} {\bibfield  {journal} {\bibinfo  {journal} {Phys.
  Rev. D}\ }\textbf {\bibinfo {volume} {83}},\ \bibinfo {pages} {124008}
  (\bibinfo {year} {2011})}\BibitemShut {NoStop}%
\bibitem [{\citenamefont {Endrizzi}\ \emph {et~al.}(2016)\citenamefont
  {Endrizzi}, \citenamefont {Ciolfi}, \citenamefont {Giacomazzo}, \citenamefont
  {Kastaun},\ and\ \citenamefont {Kawamura}}]{endrizzi2016}%
  \BibitemOpen
  \bibfield  {author} {\bibinfo {author} {\bibfnamefont {A.}~\bibnamefont
  {Endrizzi}}, \bibinfo {author} {\bibfnamefont {R.}~\bibnamefont {Ciolfi}},
  \bibinfo {author} {\bibfnamefont {B.}~\bibnamefont {Giacomazzo}}, \bibinfo
  {author} {\bibfnamefont {W.}~\bibnamefont {Kastaun}}, \ and\ \bibinfo
  {author} {\bibfnamefont {T.}~\bibnamefont {Kawamura}},\ }\href {\doibase
  10.1088/0264-9381/33/16/164001} {\bibfield  {journal} {\bibinfo  {journal}
  {Classical and Quantum Gravity}\ }\textbf {\bibinfo {volume} {33}},\ \bibinfo
  {pages} {164001} (\bibinfo {year} {2016})}\BibitemShut {NoStop}%
\bibitem [{\citenamefont {{Dietrich}}\ \emph {et~al.}(2017)\citenamefont
  {{Dietrich}}, \citenamefont {{Bernuzzi}}, \citenamefont {{Ujevic}},\ and\
  \citenamefont {{Tichy}}}]{dietrich2017a}%
  \BibitemOpen
  \bibfield  {author} {\bibinfo {author} {\bibfnamefont {T.}~\bibnamefont
  {{Dietrich}}}, \bibinfo {author} {\bibfnamefont {S.}~\bibnamefont
  {{Bernuzzi}}}, \bibinfo {author} {\bibfnamefont {M.}~\bibnamefont
  {{Ujevic}}}, \ and\ \bibinfo {author} {\bibfnamefont {W.}~\bibnamefont
  {{Tichy}}},\ }\href {\doibase 10.1103/PhysRevD.95.044045} {\bibfield
  {journal} {\bibinfo  {journal} {\prd}\ }\textbf {\bibinfo {volume} {95}},\
  \bibinfo {eid} {044045} (\bibinfo {year} {2017})},\ \Eprint
  {http://arxiv.org/abs/1611.07367} {arXiv:1611.07367 [gr-qc]} \BibitemShut
  {NoStop}%
\bibitem [{\citenamefont {Dietrich}\ \emph {et~al.}(2017)\citenamefont
  {Dietrich}, \citenamefont {Ujevic}, \citenamefont {Tichy}, \citenamefont
  {Bernuzzi},\ and\ \citenamefont {Br\"ugmann}}]{dietrich2017b}%
  \BibitemOpen
  \bibfield  {author} {\bibinfo {author} {\bibfnamefont {T.}~\bibnamefont
  {Dietrich}}, \bibinfo {author} {\bibfnamefont {M.}~\bibnamefont {Ujevic}},
  \bibinfo {author} {\bibfnamefont {W.}~\bibnamefont {Tichy}}, \bibinfo
  {author} {\bibfnamefont {S.}~\bibnamefont {Bernuzzi}}, \ and\ \bibinfo
  {author} {\bibfnamefont {B.}~\bibnamefont {Br\"ugmann}},\ }\href {\doibase
  10.1103/PhysRevD.95.024029} {\bibfield  {journal} {\bibinfo  {journal} {Phys.
  Rev. D}\ }\textbf {\bibinfo {volume} {95}},\ \bibinfo {pages} {024029}
  (\bibinfo {year} {2017})}\BibitemShut {NoStop}%
\bibitem [{\citenamefont {Constantinou}\ \emph
  {et~al.}(2015{\natexlab{a}})\citenamefont {Constantinou}, \citenamefont
  {Muccioli}, \citenamefont {Prakash},\ and\ \citenamefont
  {Lattimer}}]{Constantinou}%
  \BibitemOpen
  \bibfield  {author} {\bibinfo {author} {\bibfnamefont {C.}~\bibnamefont
  {Constantinou}}, \bibinfo {author} {\bibfnamefont {B.}~\bibnamefont
  {Muccioli}}, \bibinfo {author} {\bibfnamefont {M.}~\bibnamefont {Prakash}}, \
  and\ \bibinfo {author} {\bibfnamefont {J.~M.}\ \bibnamefont {Lattimer}},\
  }\href@noop {} {\bibfield  {journal} {\bibinfo  {journal} {Annals of
  Physics}\ }\textbf {\bibinfo {volume} {363}},\ \bibinfo {pages} {533}
  (\bibinfo {year} {2015}{\natexlab{a}})}\BibitemShut {NoStop}%
\bibitem [{\citenamefont {Wiringa}\ \emph {et~al.}(1995)\citenamefont
  {Wiringa}, \citenamefont {Stoks},\ and\ \citenamefont {Schiavilla}}]{AV18}%
  \BibitemOpen
  \bibfield  {author} {\bibinfo {author} {\bibfnamefont {R.~B.}\ \bibnamefont
  {Wiringa}}, \bibinfo {author} {\bibfnamefont {V.~G.~J.}\ \bibnamefont
  {Stoks}}, \ and\ \bibinfo {author} {\bibfnamefont {R.}~\bibnamefont
  {Schiavilla}},\ }\href@noop {} {\bibfield  {journal} {\bibinfo  {journal}
  {Phys. Rev. C}\ }\textbf {\bibinfo {volume} {51}},\ \bibinfo {pages} {38}
  (\bibinfo {year} {1995})}\BibitemShut {NoStop}%
\bibitem [{\citenamefont {Wiringa}\ and\ \citenamefont {Pieper}(2002)}]{AV6P}%
  \BibitemOpen
  \bibfield  {author} {\bibinfo {author} {\bibfnamefont {R.~B.}\ \bibnamefont
  {Wiringa}}\ and\ \bibinfo {author} {\bibfnamefont {S.~C.}\ \bibnamefont
  {Pieper}},\ }\href@noop {} {\bibfield  {journal} {\bibinfo  {journal} {Phys.
  Rev. Lett.}\ }\textbf {\bibinfo {volume} {89}},\ \bibinfo {pages} {182501}
  (\bibinfo {year} {2002})}\BibitemShut {NoStop}%
\bibitem [{\citenamefont {Pudliner}\ \emph {et~al.}(1995)\citenamefont
  {Pudliner}, \citenamefont {Pandharipande}, \citenamefont {Carlson},\ and\
  \citenamefont {Wiringa}}]{UIX_1}%
  \BibitemOpen
  \bibfield  {author} {\bibinfo {author} {\bibfnamefont {B.~S.}\ \bibnamefont
  {Pudliner}}, \bibinfo {author} {\bibfnamefont {V.~R.}\ \bibnamefont
  {Pandharipande}}, \bibinfo {author} {\bibfnamefont {J.}~\bibnamefont
  {Carlson}}, \ and\ \bibinfo {author} {\bibfnamefont {R.~B.}\ \bibnamefont
  {Wiringa}},\ }\href@noop {} {\bibfield  {journal} {\bibinfo  {journal} {Phys.
  Rev. Lett.}\ }\textbf {\bibinfo {volume} {74}},\ \bibinfo {pages} {4396}
  (\bibinfo {year} {1995})}\BibitemShut {NoStop}%
\bibitem [{\citenamefont {Carlson}\ \emph {et~al.}(1983)\citenamefont
  {Carlson}, \citenamefont {Pandharipande},\ and\ \citenamefont
  {Wiringa}}]{UIX_2}%
  \BibitemOpen
  \bibfield  {author} {\bibinfo {author} {\bibfnamefont {J.}~\bibnamefont
  {Carlson}}, \bibinfo {author} {\bibfnamefont {V.~R.}\ \bibnamefont
  {Pandharipande}}, \ and\ \bibinfo {author} {\bibfnamefont {R.~B.}\
  \bibnamefont {Wiringa}},\ }\href@noop {} {\bibfield  {journal} {\bibinfo
  {journal} {Nucl. Phys. A}\ }\textbf {\bibinfo {volume} {401}},\ \bibinfo
  {pages} {59} (\bibinfo {year} {1983})}\BibitemShut {NoStop}%
\bibitem [{\citenamefont {Lovato}\ \emph {et~al.}(2022)\citenamefont {Lovato},
  \citenamefont {Bombaci}, \citenamefont {Logoteta}, \citenamefont {Piarulli},\
  and\ \citenamefont {Wiringa}}]{Lovato:2022}%
  \BibitemOpen
  \bibfield  {author} {\bibinfo {author} {\bibfnamefont {A.}~\bibnamefont
  {Lovato}}, \bibinfo {author} {\bibfnamefont {I.}~\bibnamefont {Bombaci}},
  \bibinfo {author} {\bibfnamefont {D.}~\bibnamefont {Logoteta}}, \bibinfo
  {author} {\bibfnamefont {M.}~\bibnamefont {Piarulli}}, \ and\ \bibinfo
  {author} {\bibfnamefont {R.~B.}\ \bibnamefont {Wiringa}},\ }\href@noop {} {}
  (\bibinfo {year} {2022})\BibitemShut {NoStop}%
\bibitem [{\citenamefont {Akmal}\ \emph
  {et~al.}(1998{\natexlab{a}})\citenamefont {Akmal}, \citenamefont
  {Pandharipande},\ and\ \citenamefont {Ravenhall}}]{Akmal:1997}%
  \BibitemOpen
  \bibfield  {author} {\bibinfo {author} {\bibfnamefont {A.}~\bibnamefont
  {Akmal}}, \bibinfo {author} {\bibfnamefont {V.~R.}\ \bibnamefont
  {Pandharipande}}, \ and\ \bibinfo {author} {\bibfnamefont {D.~G.}\
  \bibnamefont {Ravenhall}},\ }\href@noop {} {\bibfield  {journal} {\bibinfo
  {journal} {Phys. Rev. C}\ }\textbf {\bibinfo {volume} {58}},\ \bibinfo
  {pages} {1804} (\bibinfo {year} {1998}{\natexlab{a}})}\BibitemShut {NoStop}%
\bibitem [{\citenamefont {Akmal}\ \emph
  {et~al.}(1998{\natexlab{b}})\citenamefont {Akmal}, \citenamefont
  {Pandharipande},\ and\ \citenamefont {Ravenhall}}]{Akmal:1998}%
  \BibitemOpen
  \bibfield  {author} {\bibinfo {author} {\bibfnamefont {A.}~\bibnamefont
  {Akmal}}, \bibinfo {author} {\bibfnamefont {V.~R.}\ \bibnamefont
  {Pandharipande}}, \ and\ \bibinfo {author} {\bibfnamefont {D.~G.}\
  \bibnamefont {Ravenhall}},\ }\href@noop {} {\bibfield  {journal} {\bibinfo
  {journal} {Phys. Rev. C}\ }\textbf {\bibinfo {volume} {58}},\ \bibinfo
  {pages} {1804} (\bibinfo {year} {1998}{\natexlab{b}})}\BibitemShut {NoStop}%
\bibitem [{\citenamefont {Lovato}\ \emph {et~al.}(2013)\citenamefont {Lovato},
  \citenamefont {Losa},\ and\ \citenamefont {Benhar}}]{lovato2013}%
  \BibitemOpen
  \bibfield  {author} {\bibinfo {author} {\bibfnamefont {A.}~\bibnamefont
  {Lovato}}, \bibinfo {author} {\bibfnamefont {C.}~\bibnamefont {Losa}}, \ and\
  \bibinfo {author} {\bibfnamefont {O.}~\bibnamefont {Benhar}},\ }\href
  {\doibase https://doi.org/10.1016/j.nuclphysa.2013.01.029} {\bibfield
  {journal} {\bibinfo  {journal} {Nuclear Physics A}\ }\textbf {\bibinfo
  {volume} {901}},\ \bibinfo {pages} {22} (\bibinfo {year} {2013})}\BibitemShut
  {NoStop}%
\bibitem [{\citenamefont {Benhar}\ and\ \citenamefont
  {Lovato}(2017{\natexlab{b}})}]{eos0}%
  \BibitemOpen
  \bibfield  {author} {\bibinfo {author} {\bibfnamefont {O.}~\bibnamefont
  {Benhar}}\ and\ \bibinfo {author} {\bibfnamefont {A.}~\bibnamefont
  {Lovato}},\ }\href@noop {} {\bibfield  {journal} {\bibinfo  {journal} {Phys.
  Rev. C}\ }\textbf {\bibinfo {volume} {96}},\ \bibinfo {pages} {054301}
  (\bibinfo {year} {2017}{\natexlab{b}})}\BibitemShut {NoStop}%
\bibitem [{\citenamefont {Benhar}\ and\ \citenamefont
  {Valli}(2007)}]{benharvalli2007}%
  \BibitemOpen
  \bibfield  {author} {\bibinfo {author} {\bibfnamefont {O.}~\bibnamefont
  {Benhar}}\ and\ \bibinfo {author} {\bibfnamefont {M.}~\bibnamefont {Valli}},\
  }\href {\doibase 10.1103/PhysRevLett.99.232501} {\bibfield  {journal}
  {\bibinfo  {journal} {Phys. Rev. Lett.}\ }\textbf {\bibinfo {volume} {99}},\
  \bibinfo {pages} {232501} (\bibinfo {year} {2007})}\BibitemShut {NoStop}%
\bibitem [{\citenamefont {Alford}\ \emph {et~al.}(2021)\citenamefont {Alford},
  \citenamefont {Harutyunyan},\ and\ \citenamefont {Sedrakian}}]{alford2021}%
  \BibitemOpen
  \bibfield  {author} {\bibinfo {author} {\bibfnamefont {M.}~\bibnamefont
  {Alford}}, \bibinfo {author} {\bibfnamefont {A.}~\bibnamefont {Harutyunyan}},
  \ and\ \bibinfo {author} {\bibfnamefont {A.}~\bibnamefont {Sedrakian}},\
  }\href {\doibase 10.1103/PhysRevD.104.103027} {\bibfield  {journal} {\bibinfo
   {journal} {Phys. Rev. D}\ }\textbf {\bibinfo {volume} {104}},\ \bibinfo
  {pages} {103027} (\bibinfo {year} {2021})}\BibitemShut {NoStop}%
\bibitem [{\citenamefont {Baldo}\ and\ \citenamefont
  {Ferreira}(1999)}]{baldo1999}%
  \BibitemOpen
  \bibfield  {author} {\bibinfo {author} {\bibfnamefont {M.}~\bibnamefont
  {Baldo}}\ and\ \bibinfo {author} {\bibfnamefont {L.~S.}\ \bibnamefont
  {Ferreira}},\ }\href {\doibase 10.1103/PhysRevC.59.682} {\bibfield  {journal}
  {\bibinfo  {journal} {Phys. Rev. C}\ }\textbf {\bibinfo {volume} {59}},\
  \bibinfo {pages} {682} (\bibinfo {year} {1999})}\BibitemShut {NoStop}%
\bibitem [{\citenamefont {Burgio}\ and\ \citenamefont {Schulze}(2010)}]{BS:AA}%
  \BibitemOpen
  \bibfield  {author} {\bibinfo {author} {\bibfnamefont {G.~F.}\ \bibnamefont
  {Burgio}}\ and\ \bibinfo {author} {\bibfnamefont {H.-J.}\ \bibnamefont
  {Schulze}},\ }\href@noop {} {\bibfield  {journal} {\bibinfo  {journal}
  {Astronomy \& Astrophysics}\ }\textbf {\bibinfo {volume} {518}},\ \bibinfo
  {pages} {A10} (\bibinfo {year} {2010})}\BibitemShut {NoStop}%
\bibitem [{\citenamefont {Constantinou}\ \emph
  {et~al.}(2015{\natexlab{b}})\citenamefont {Constantinou}, \citenamefont
  {Muccioli}, \citenamefont {Prakash},\ and\ \citenamefont
  {Lattimer}}]{constantinou2015}%
  \BibitemOpen
  \bibfield  {author} {\bibinfo {author} {\bibfnamefont {C.}~\bibnamefont
  {Constantinou}}, \bibinfo {author} {\bibfnamefont {B.}~\bibnamefont
  {Muccioli}}, \bibinfo {author} {\bibfnamefont {M.}~\bibnamefont {Prakash}}, \
  and\ \bibinfo {author} {\bibfnamefont {J.~M.}\ \bibnamefont {Lattimer}},\
  }\href {\doibase https://doi.org/10.1016/j.aop.2015.10.003} {\bibfield
  {journal} {\bibinfo  {journal} {Annals of Physics}\ }\textbf {\bibinfo
  {volume} {363}},\ \bibinfo {pages} {533} (\bibinfo {year}
  {2015}{\natexlab{b}})}\BibitemShut {NoStop}%
\bibitem [{\citenamefont {Raithel}\ \emph
  {et~al.}(2021{\natexlab{b}})\citenamefont {Raithel}, \citenamefont {Özel},\
  and\ \citenamefont {Psaltis}}]{raithel2019erratum}%
  \BibitemOpen
  \bibfield  {author} {\bibinfo {author} {\bibfnamefont {C.~A.}\ \bibnamefont
  {Raithel}}, \bibinfo {author} {\bibfnamefont {F.}~\bibnamefont {Özel}}, \
  and\ \bibinfo {author} {\bibfnamefont {D.}~\bibnamefont {Psaltis}},\ }\href
  {\doibase 10.3847/1538-4357/ac0630} {\bibfield  {journal} {\bibinfo
  {journal} {The Astrophysical Journal}\ }\textbf {\bibinfo {volume} {915}},\
  \bibinfo {pages} {73} (\bibinfo {year} {2021}{\natexlab{b}})}\BibitemShut
  {NoStop}%
\bibitem [{\citenamefont {Camelio}\ \emph {et~al.}(2021)\citenamefont
  {Camelio}, \citenamefont {Dietrich}, \citenamefont {Rosswog},\ and\
  \citenamefont {Haskell}}]{camelio2021}%
  \BibitemOpen
  \bibfield  {author} {\bibinfo {author} {\bibfnamefont {G.}~\bibnamefont
  {Camelio}}, \bibinfo {author} {\bibfnamefont {T.}~\bibnamefont {Dietrich}},
  \bibinfo {author} {\bibfnamefont {S.}~\bibnamefont {Rosswog}}, \ and\
  \bibinfo {author} {\bibfnamefont {B.}~\bibnamefont {Haskell}},\ }\href
  {\doibase 10.1103/PhysRevD.103.063014} {\bibfield  {journal} {\bibinfo
  {journal} {Phys. Rev. D}\ }\textbf {\bibinfo {volume} {103}},\ \bibinfo
  {pages} {063014} (\bibinfo {year} {2021})}\BibitemShut {NoStop}%
\bibitem [{\citenamefont {Most}\ \emph {et~al.}(2021)\citenamefont {Most},
  \citenamefont {Harris}, \citenamefont {Plumberg}, \citenamefont {Alford},
  \citenamefont {Noronha}, \citenamefont {Noronha-Hostler}, \citenamefont
  {Pretorius}, \citenamefont {Witek},\ and\ \citenamefont
  {Yunes}}]{mostetal2021}%
  \BibitemOpen
  \bibfield  {author} {\bibinfo {author} {\bibfnamefont {E.~R.}\ \bibnamefont
  {Most}}, \bibinfo {author} {\bibfnamefont {S.~P.}\ \bibnamefont {Harris}},
  \bibinfo {author} {\bibfnamefont {C.}~\bibnamefont {Plumberg}}, \bibinfo
  {author} {\bibfnamefont {M.~G.}\ \bibnamefont {Alford}}, \bibinfo {author}
  {\bibfnamefont {J.}~\bibnamefont {Noronha}}, \bibinfo {author} {\bibfnamefont
  {J.}~\bibnamefont {Noronha-Hostler}}, \bibinfo {author} {\bibfnamefont
  {F.}~\bibnamefont {Pretorius}}, \bibinfo {author} {\bibfnamefont
  {H.}~\bibnamefont {Witek}}, \ and\ \bibinfo {author} {\bibfnamefont
  {N.}~\bibnamefont {Yunes}},\ }\href {\doibase 10.1093/mnras/stab2793}
  {\bibfield  {journal} {\bibinfo  {journal} {Monthly Notices of the Royal
  Astronomical Society}\ }\textbf {\bibinfo {volume} {509}},\ \bibinfo {pages}
  {1096} (\bibinfo {year} {2021})},\ \Eprint
  {http://arxiv.org/abs/https://academic.oup.com/mnras/article-pdf/509/1/1096/41135859/stab2793.pdf}
  {https://academic.oup.com/mnras/article-pdf/509/1/1096/41135859/stab2793.pdf}
  \BibitemShut {NoStop}%
\bibitem [{\citenamefont {Most}\ \emph {et~al.}(2022)\citenamefont {Most},
  \citenamefont {Haber}, \citenamefont {Harris}, \citenamefont {Zhang},
  \citenamefont {Alford},\ and\ \citenamefont {Noronha}}]{mostetal2022}%
  \BibitemOpen
  \bibfield  {author} {\bibinfo {author} {\bibfnamefont {E.~R.}\ \bibnamefont
  {Most}}, \bibinfo {author} {\bibfnamefont {A.}~\bibnamefont {Haber}},
  \bibinfo {author} {\bibfnamefont {S.~P.}\ \bibnamefont {Harris}}, \bibinfo
  {author} {\bibfnamefont {Z.}~\bibnamefont {Zhang}}, \bibinfo {author}
  {\bibfnamefont {M.~G.}\ \bibnamefont {Alford}}, \ and\ \bibinfo {author}
  {\bibfnamefont {J.}~\bibnamefont {Noronha}},\ }\href {\doibase
  10.48550/ARXIV.2207.00442} {\enquote {\bibinfo {title} {Emergence of
  microphysical viscosity in binary neutron star post-merger dynamics},}\ }
  (\bibinfo {year} {2022})\BibitemShut {NoStop}%
\bibitem [{\citenamefont {Alford}\ and\ \citenamefont
  {Harris}(2018)}]{alford2018beta}%
  \BibitemOpen
  \bibfield  {author} {\bibinfo {author} {\bibfnamefont {M.~G.}\ \bibnamefont
  {Alford}}\ and\ \bibinfo {author} {\bibfnamefont {S.~P.}\ \bibnamefont
  {Harris}},\ }\href {\doibase 10.1103/PhysRevC.98.065806} {\bibfield
  {journal} {\bibinfo  {journal} {Phys. Rev. C}\ }\textbf {\bibinfo {volume}
  {98}},\ \bibinfo {pages} {065806} (\bibinfo {year} {2018})}\BibitemShut
  {NoStop}%
\end{thebibliography}
